\documentclass[aps,twocolumn, superscriptaddress, 10pt,floatfix,showpacs,prl]{revtex4-1}

\usepackage[normalem]{ulem}

\usepackage{graphicx}
\usepackage{color}
\usepackage{amsfonts, amsmath, amsthm, amscd,amssymb}

\usepackage{tensor}
\usepackage{xfrac}

\def \der {\mathrm{d}}

\def\epp{\: .}
\def\epc{\: ,}

\newcommand{\ket}[1]{{\left| #1 \right\rangle}}
\newcommand{\bra}[1]{{\left\langle #1 \right|}}

\newcommand{\matrixel}[3]{\left\langle #1 \vphantom{#2#3} \right| #2 \left| #3 \vphantom{#1#2} \right\rangle}

\newcommand{\smatrixel}[3]{\langle #1 \vphantom{#2#3} | #2 | #3 \vphantom{#1#2} \rangle}

\def\rhosp{\rho^\text{sp}}

\def\limth{\lim\nolimits_\text{th}}

\def\lam{\lambda}

\def\blam{{\boldsymbol{\lambda}}}

\def\boldx{{\boldsymbol{x}}}
\def\bx{{\boldsymbol{x}}}

\def\b0{\boldsymbol{0}}
\def\bmu{\boldsymbol{\mu}}

\def\hH{{\hat{H}}}

\def\hQ{{\hat{Q}}}

\def\hPsi{{\hat{\Psi}}}
\def\hrho{{\hat{\rho}}}

\def\hU{{\hat{U}}}

\def\GS{\text{GS}}

\begin{document}

\title{Separation of Timescales in a Quantum Newton's Cradle}

\author{R. van den Berg}
\email{R.vandenBerg2@uva.nl}
\affiliation{Institute for Theoretical Physics, University of Amsterdam, 
Science Park 904,\\ 1098 XH Amsterdam, The Netherlands}

\author{B. Wouters}
\affiliation{Institute for Theoretical Physics, University of Amsterdam, 
Science Park 904,\\ 1098 XH Amsterdam, The Netherlands}

\author{S. Eli\"ens}
\affiliation{Institute for Theoretical Physics, University of Amsterdam, 
Science Park 904,\\ 1098 XH Amsterdam, The Netherlands}

\author{J. De Nardis}
\affiliation{Institute for Theoretical Physics, University of Amsterdam, 
Science Park 904,\\ 1098 XH Amsterdam, The Netherlands}

\author{R.M. Konik}
\affiliation{CMPMS Dept. Bldg 734 Brookhaven National Laboratory, Upton NY 11973, USA}

\author{J.-S. Caux}
\affiliation{Institute for Theoretical Physics, University of Amsterdam, 
Science Park 904,\\ 1098 XH Amsterdam, The Netherlands}
\date{\today}

\begin{abstract}For strongly repulsive bosons in one dimension, we provide detailed modeling of the Bragg pulse used in quantum Newton's cradle-like settings or in Bragg spectroscopy experiments. By employing the Fermi-Bose mapping for a finite harmonically trapped gas and the Quench Action approach for a thermodynamic system on a ring, we reconstruct the exact post-pulse many-body time evolution of Lieb-Liniger gases in the Tonks-Girardeau limit, together with their changing local density profile and momentum distribution. Our results display a clear separation of timescales between rapid and trap-insensitive relaxation immediately after the pulse, followed by slow in-trap periodic behaviour. 
\end{abstract}

\maketitle

The study of many-body quantum physics has in recent years been transformed by the progress achieved in experiments on ultracold atoms \cite{2008_Bloch_RMP_80}. The context of one-dimensional (1D) bosonic gases in particular provides a fertile ground for investigating physics beyond traditional paradigms \cite{2011_Cazalilla_RMP_83}, with concepts such as the Luttinger liquid and exact solvability \cite{GiamarchiBOOK} playing a primary role.

One of the main experimental probes of cold gases is Bragg
spectroscopy \cite{1988_Martin_PRL_60,1999_Stenger_PRL_82,2005_Ozeri_RMP_77}, which
consists in applying a pulsed monochromatic laser grating onto the
gas, thereby creating excitations at (multiples of) the recoil momentum $q$, where $q/2$ is the wavevector of the laser. The precise time-dependent form of the pulse can be chosen to optimize (de)population of specific quantum states. In \cite{2005_Wang_PRL_94,2005_Wu_PRA_71}, a two-pulse sequence was used to optimize the population of the first $\pm q$ satellites as compared to the zero-momentum ground state of a Bose-Einstein condensate. The theoretical description of such a sequence relied on a two-state model in which many-body dynamics are not included. In one dimension however, many-body effects are inescapable. One of the fundamental models in this context is the Lieb-Liniger gas \cite{1963_Lieb_PR_130} of bosons in a 1D continuum interacting with a $\delta$-function potential, giving a proper description of atoms in tight transverse confinement \cite{1998_Olshanii_PRL_81}. This model is relevant to the description of experiments, most prominently the famous quantum Newton's cradle experiment \cite{2006_Kinoshita_NATURE_440}, in which a Bragg pulse is used to initiate the oscillations. Bragg spectroscopy has also recently been used to investigate correlated 1D Bose gases of rubidium \cite{2015_Fabbri_PRA_91} and cesium \cite{2015_Meinert_PRL_115}. In these, the heating of the gas resulting from the Bragg pulse was measured and matched using linear response to theoretical calculations of the dynamical structure factor of the Lieb-Liniger gas at finite temperature \cite{2014_Panfil_PRA_89}. 

Our main objective here is to model the effects of Bragg pulses theoretically for correlated 1D Bose gases, from first principles and without approximation (and thus beyond linear response), for experimentally relevant setups. We focus on the Tonks-Girardeau limit \cite{1936_Tonks_PR_50, 1960_Girardeau_JMP_1,2009_Haller_SCIENCE_325} of strongly repulsive Lieb-Liniger bosons both on a periodic interval and in confining traps \cite{2005_Minguzzi_PRL_94,2007_Pezer_PRL_98,2008_Gangardt_PRA_77,
2010_Muth_NJP_12,2011_Schenke_PRA_84,2013_Astrakharchik_EPL_102,
2013_Collura_PRL_110,2014_Quinn_PRA_90,2015_Cartarius_arXiv}. Instantaneous Bragg pulses of varying amplitude $A$ and wavevector $q$ are studied via their effect on physical observables: the time-dependent local density of the gas, and the experimentally more easily accessible momentum distribution function (MDF).

We start by modeling the Bragg pulse as a standing wave forming a one-body potential $V(x)= V_0 \cos (qx)$ that couples to the density $\hrho(x) = \hPsi^\dag(x) \hPsi(x)$, where the Bose fields obey the canonical equal-time commutation relations, $\big[ \hPsi(x),\, \hPsi^\dag(y)\big] = \delta(x-y)$. 
For a general Bragg pulse the gas is perturbed for a finite duration $T_0$.  We will however consider the regime where the motion of the particles during the pulse can be neglected (the Raman-Nath limit), in which case the Bragg pulse is also referred to as a Kapitza-Dirac pulse ~\cite{1933_Kapitza_Proc,2001_Freimund_Nat}. Taking the limit $T_0 \rightarrow 0$ such that $A = V_0 T_0$ is kept finite,
the Bragg pulse operator $\hU_B$ is given by
\begin{align}
\hat{U}_B (q,A) =  \exp \Bigg( - i A \int dx \, \cos(qx) \hPsi^{\dagger}(x)\hPsi(x) \Bigg),
\label{eq:O_bragg}
\end{align}
where we have used the convention $\hbar=1$. 
Applying such an instantaneous Bragg pulse to a ground state $\ket{\psi_{\text{GS}}}$ yields a post-pulse state $\ket{\psi_{q,A}} = \hU_B (q,A) \ket{\psi_{\text{GS}}}$, which can be interpreted as the initial state of a quantum quench \cite{2006_Calabrese_PRL_96,2007_Rigol_PRL_98,2011_Polkovnikov_RMP_83}. Typical experimental pulses~\cite{2006_Kinoshita_NATURE_440,2009_Sapiro_PRA_79,2015_Fabbri_PRA_91,2015_Meinert_PRL_115} correspond to Bragg momentum $q \thicksim 2 \pi n$ and $A \thicksim 1$, where $n$ is the mean density.

The post-pulse time evolution is driven by the Lieb-Liniger (LL) model of interacting bosons 
\begin{align}
H_{\text{LL}} =  &-\sum_{i=1}^N \frac{1}{2m}\frac{\partial^2}{\partial x_i^2}  + 2c \sum_{1\leq i < j \leq N} \delta(x_i - x_j).
\label{eq:LL_Ham_1st_quantization}
\end{align}
In what follows we will focus on the hard-core Tonks-Girardeau (TG) limit $c \to \infty$ \cite{1936_Tonks_PR_50, 1960_Girardeau_JMP_1} and we will consider the model both on a ring (periodic boundary conditions) and in a parabolic trapping potential. 

In the hard-core limit, the bosonic many-body wavefunction can be related through the Fermi-Bose (FB) mapping~\cite{1960_Girardeau_JMP_1} to the many-body wavefunction of free fermions $\psi_B(\bx;t) = \prod_{1\leq i<j \leq N} \mathrm{sgn}(x_i-x_j) \psi_F(\bx;t)$, where \smash{$\bx=\{x_j\}_{j=1}^N$} and the fermionic wavefunction is the usual Slater determinant of the free single-particle (SP) wavefunctions, $\psi_F(\bx;t) = \text{det}_N \left[\psi_{j}(x_i;t)\right]/\sqrt{N!}$.  Following \cite{2007_Pezer_PRL_98,2015_Cartarius_arXiv,2013_Collura_PRL_110}, 
the bosonic one-body density matrix, defined as $\langle \hPsi^{\dagger}(x,t) \hPsi(y,t)\rangle = N \! \int \! \der x_2 ... \der x_N \, \left.\psi^*_B(\bx,t)\right|_{x_1=x} \left.\psi_B(\bx,t)\right|_{x_1=y}$, is given in terms of a single determinant involving the time-dependent fermionic SP states. This allows for an efficient computation of the MDF $\langle \hat{n}(k,t)\rangle = \frac{1}{2\pi}\int \der x \,\der y \;e^{i(x-y)k}\langle \hPsi^{\dagger}(x,t) \hPsi(y,t)\rangle$.

Starting with the ring geometry, our ground state consists of SP plane waves, on which the Bragg pulse imprints a cosine phase due to the one-body potential,
\begin{align}
\psi_j(x;0) = \frac{1}{\sqrt{L}}e^{-i A \cos(qx)} e^{-i \lambda^{\text{GS}}_j x},
\label{eq:psi_0_ring}
\end{align}
with ground-state rapidities $\{  \lambda^{\text{GS}}_j = \tfrac{2\pi}{L}\left( -\frac{N+1}{2} + j\right)\}_{j=1}^N $ forming a Fermi sea with Fermi momentum $\lam_F = \lam_N^{\text{GS}}$. 
Note that the Bragg momentum is quantized due to the periodic boundary conditions of the ring, $q=\tfrac{2\pi}{L}n_q$ with $ n_q \in \mathbb{N}$. 
Expanding Eq. \eqref{eq:psi_0_ring} in plane waves, the time-dependent SP wavefunctions after the Bragg pulse yield
\begin{align}
\psi_j(x;t)  = \sum_{\beta=-\infty}^{\infty} I_{\beta}(-iA) \frac{1}{\sqrt{L}} e^{-i (\lambda_j +\beta q) x}  e^{-i (\lambda_j+\beta q)^2 t/2m},
\end{align}
where $I_{\beta}(z)$ is the modified Bessel function of the first kind.

Contrary to the finite-size Fermi-Bose mapping, the Generalized Gibbs Ensemble (GGE)~\cite{2007_Rigol_PRL_98,2008_Rigol_NATURE_452} and the Quench Action (QA) approach~\cite{2013_Caux_PRL_110,2014_DeNardis_PRA_89} enable the study of the Bragg pulsed system (on a ring) in the thermodynamic limit ($N \to \infty$ with $N/L$ fixed). The GGE can be constructed using the infinite number of conserved charges $\{ \hQ_{\alpha} \}_{\alpha=1}^\infty$ provided by the integrability of the LL model, with $\hQ_2 = 2m\hH$, and eigenvalues $Q_{\alpha} (\blam) = \sum_{j=1}^N \lam_j^{\alpha}$ associated to a Bethe state $\ket{\blam}= \ket{\lambda_1,...,\lambda_N}$. The expectation values of the charges on the initial post-pulse state can be computed using the overlaps $\langle \blam | \psi_{q,A} \rangle$, which can be derived from the matrix elements for the Bragg pulse between two Bethe states $\ket{\blam}$ and $\ket{\bmu}$ \cite{SupMat}
\begin{equation} \label{eq:matrix_elements}
\frac{\smatrixel{\bmu}{\hU_B (q,A)}{\blam}}{L^N} = \text{det}_N\! \left[  
I_{\frac{\lam_j-\mu_k}{q}}(-iA)\, \delta^{(q)}_{\lam_j,\mu_k} 
 \right]  ,
\end{equation}
where we defined $\delta^{(q)}_{\lambda,\mu} = \delta_{(\lambda-\mu)\mathrm{mod}\; q,0}$.  Taking the thermodynamic limit, the energy density of the system after the Bragg pulse yields $\limth \frac1L \left\langle \psi_{q,A} \right| \hQ_2 \left| \psi_{q,A} \right\rangle  = \tfrac13 \lam_F^2 + \tfrac12 (q A)^2$.
The GGE logic~\cite{2007_Rigol_PRL_98,2008_Rigol_NATURE_452} then requires the expectation values of all charges to be reproduced by the equilibrated post-pulse system, described by a density of rapidites  $\rhosp_{q,A}(\lambda)$, i.e.
\begin{equation} \label{eq:equilibrium_expectation_value_QA}
\limth \frac1L \left\langle \psi_{q,A} \right| \hQ_{\alpha} \left| \psi_{q,A} \right\rangle =  \int_{-\infty}^\infty d\lam\, \rhosp_{q,A} (\lam) \, \lam^{\alpha} 
\epc
\end{equation}
for all $\alpha \in \mathbb{N}$. This leads to the following stationary-state distribution \cite{SupMat,2011_Bevilacqua_JMP_52}, 
\begin{equation} \label{eq:saddle_point_bragg_pulse}
\rhosp_{q,A}(\lam) = \frac{1}{2\pi} \! \sum_{\beta\in\mathbb{Z}} \big[ \theta(\lam-\beta q + \lam_F) - \theta(\lam-\beta q - \lam_F) \big] \!  \left| I_{\beta} ( iA) \right|^2 
\end{equation}
where $\theta$ is the Heaviside step function.
The saddle point distribution is a  sum of copies of the ground-state density of rapidities, $\rho_{\text{GS}}(\lambda) = \frac{1}{2\pi} \big[\theta(\lam +\lam_F) - \theta(\lam - \lam_F)\big] $, shifted by multiples of $q$ and weighted by the modified Bessel functions. 

\begin{figure}[ht]
	\includegraphics[width=0.99 \columnwidth]{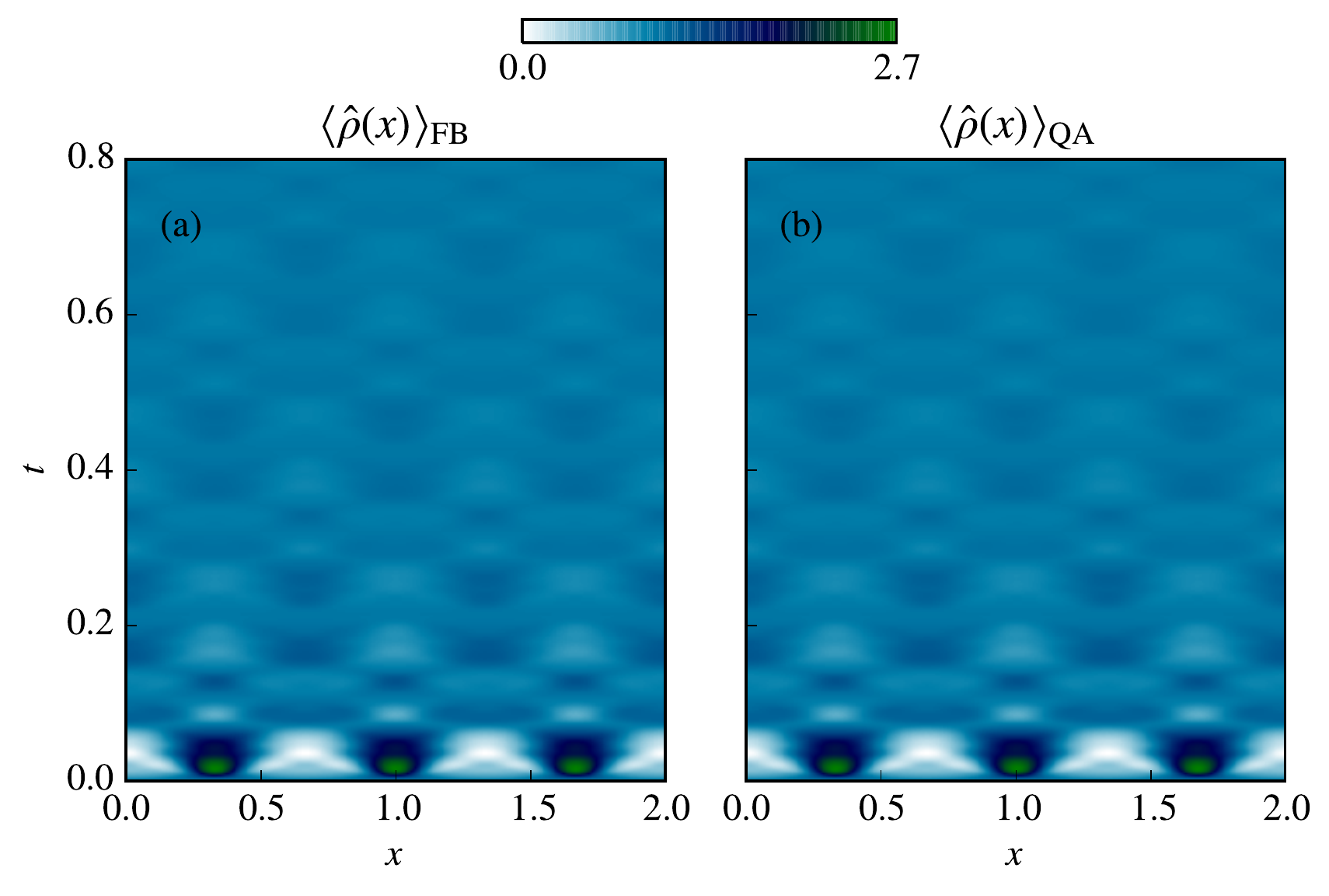}
		\caption{ Time evolution of the density after a Bragg pulse with $q=3 \pi$ and $A=1.5$, computed by (a) the FB mapping and (b) the QA approach. The relative differences between the two results due to finite-size effects are less than $0.4 \%$.}
	\label{fig:density_QA_Fermibose}
\end{figure}

This form of the stationary state is consistent with the QA approach \cite{SupMat}, which furthermore states that the time evolution of local observables after a quantum quench is given by a sum over particle-hole excitations in the vicinity of  $\rhosp_{q,A}(\lambda)$~\cite{2013_Caux_PRL_110, 2014_DeNardis_PRA_89,DeNardisg2TimeEv}. One easily obtains the time evolution of the density of the gas,
\begin{align} \label{eq:time_evolution_density}
 \limth &\matrixel{ \psi_{q,A} (t)}{ \hat{\rho}(x)  }{ \psi_{q,A} (t) }  
 =  \frac{n m}{ q  \lam_F t} \times \\
 &\sum_{\beta=-\infty}^{\infty} J_{\beta }(-2A\sin(q^2\beta t/2m)) \cos(xq\beta) \frac{\sin(q \lam_F  \beta t/m )}{\beta } \notag \epc
\end{align}
with $J_\beta(z)$ the Bessel function of the first kind. The time evolution of the density is compared to the FB result for $N=50$ in Fig.~\ref{fig:density_QA_Fermibose} and shows excellent agreement, with relative differences of order $0.4\%$ due to finite-size effects in the FB calculations. Throughout the paper, all data is produced setting $m=1$. As a consequence of the Raman-Nath limit, the initial post-pulse density (at $t=0$) is unaltered from the flat ground state profile. 
A sharp density profile then develops, mimicking the one-body cosine potential that was instantaneously turned on and off,
 followed by relaxation back to a flat profile at time scales $t \thicksim m/q\lambda_F = (qv_s)^{-1}$, with $v_s = \pi n/m$ the sound velocity. 

\begin{figure}[ht]
	\includegraphics[width=0.99 \columnwidth]{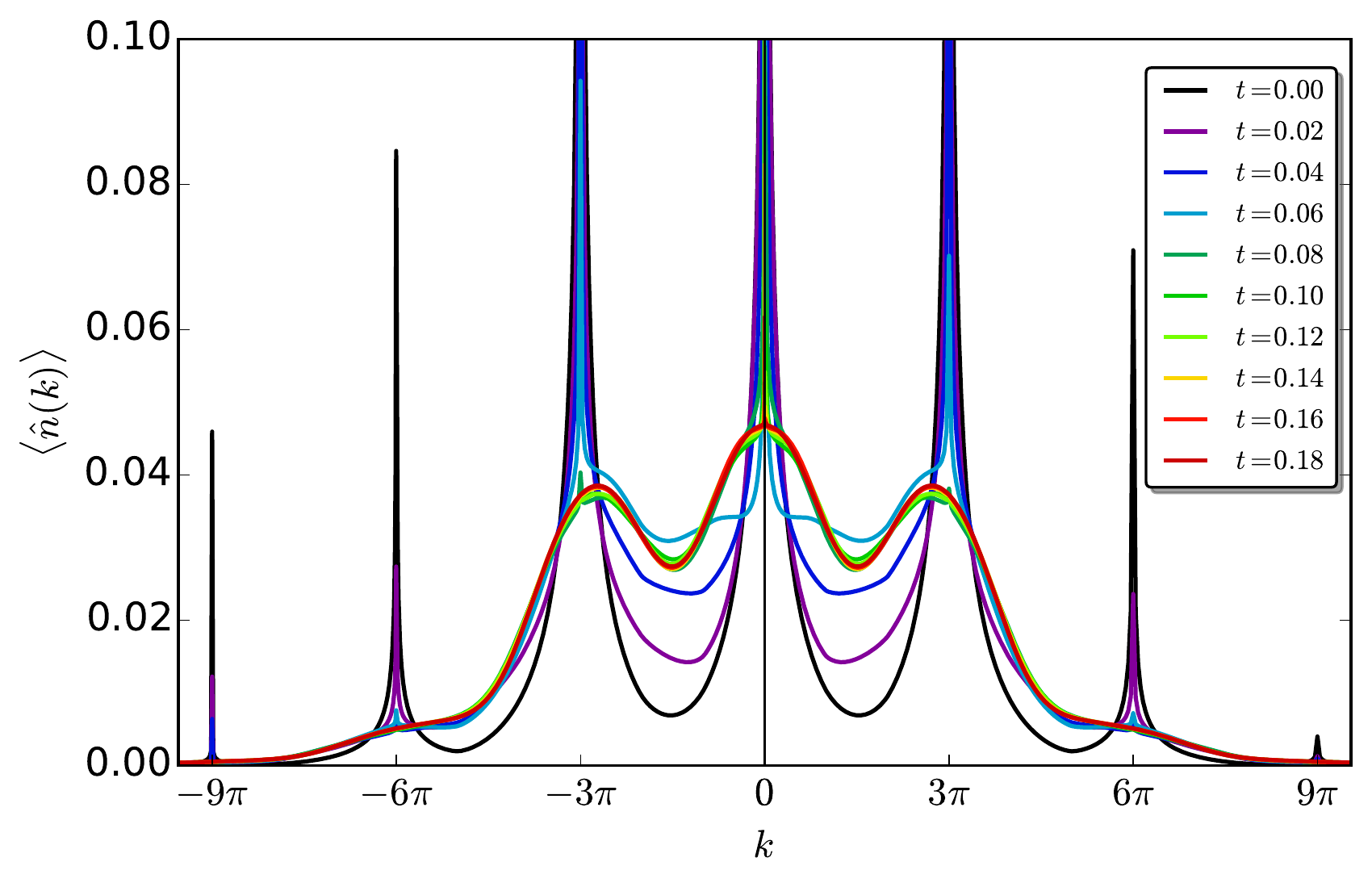}
		\caption{Time evolution of the MDF after a Bragg pulse with $q=3 \pi$ and $A=1.4$, computed with the QA approach (left half) and the FB mapping (right half). Because the FB mapping treats a finite system ($N=50$) the momenta are quantized, causing less pronounced peaks for short times. All other results are in excellent agreement with the QA computations.}
	\label{fig:TG_mom_ring_QA_Fermibose}
\end{figure}

The QA approach also provides access to the time evolution of the MDF~\cite{2014_Nardis_JSTAT_P12012,SupMat}. 
The result is plotted in Fig.~\ref{fig:TG_mom_ring_QA_Fermibose} along with the FB result for $N=50$. Except for minor disagreements in the sharp peaks due to finite-size effects, the large-system-size dynamics after the Bragg pulse is again well captured by a Fermi-Bose mapping for $N=50$ particles. At $t=0$, using the commutation relations between the Bragg operator and the Bose fields, one can show that the momentum distribution is simply a sum of copies of the ground-state MDF \cite{SupMat}, with a small-$k$ divergence $\langle \hat{n}(k) \rangle_{\text{GS}} \thicksim k^{-1/2}$, centered around multiples of $q$. The sharply peaked MDF then relaxes to a characteristic ghost-like shape \cite{2006_Kinoshita_NATURE_440}, with the mixing of particles with different momenta causing a substantially broadened stationary MDF.

\begin{figure}[ht]
	\includegraphics[width=0.99 \columnwidth]{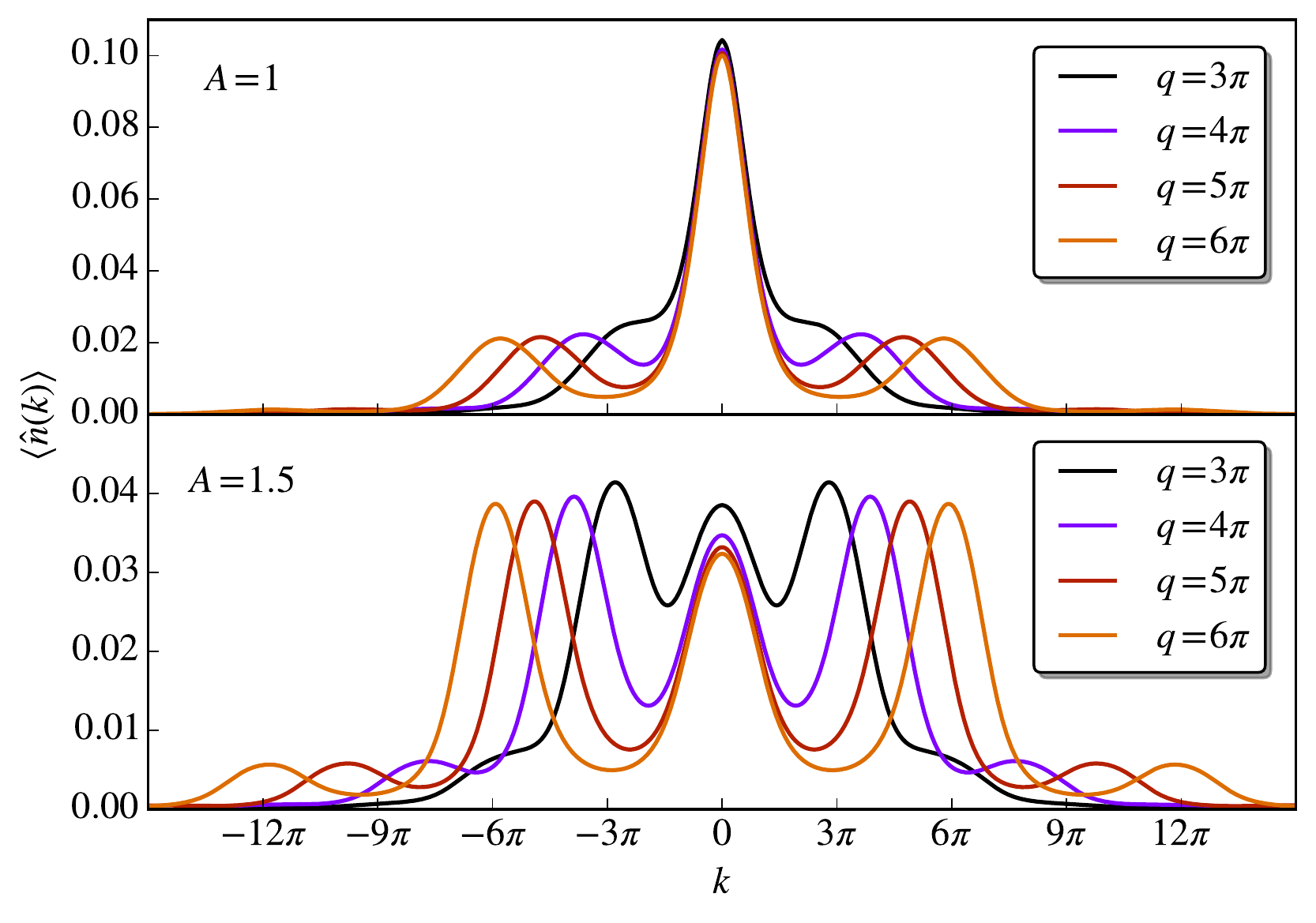}
		\caption{The relaxed MDF function for $N=50$ particles as a function of the Bragg momentum $q$. In the upper panel $A=1$, and in the lower panel $A$ is fixed to $1.5$. }
	\label{fig:FB_relaxed}
\end{figure}

In Fig. \ref{fig:FB_relaxed} the equilibrated MDF is shown for different values of $q$ and $A$. Similar to the initial MDF, the late-time distribution behaves like a superposition of independent peaks shifted to multiples of $q$. The width of each satellite shows no dependence on the value of $q$, and is only influenced by the choice of $A$. 
Since in the limit of $A\rightarrow 0$ the resulting MDF is just that of the ground state and would stay constant as time progresses, the broadening can be ascribed to interactions between particles belonging to different copies of the ground state density of rapidities in $\rhosp_{q,A}$.

\begin{figure}[ht]
	\begin{center}$
	\begin{array}[b]{c}
	\includegraphics[width=1.\columnwidth]{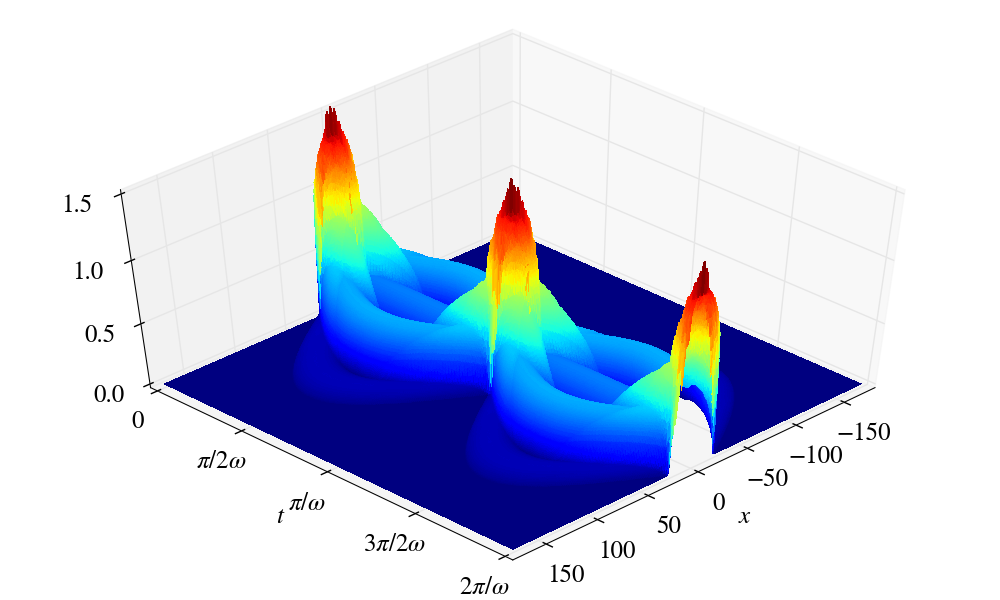}\\
	\includegraphics[width=1.\columnwidth]{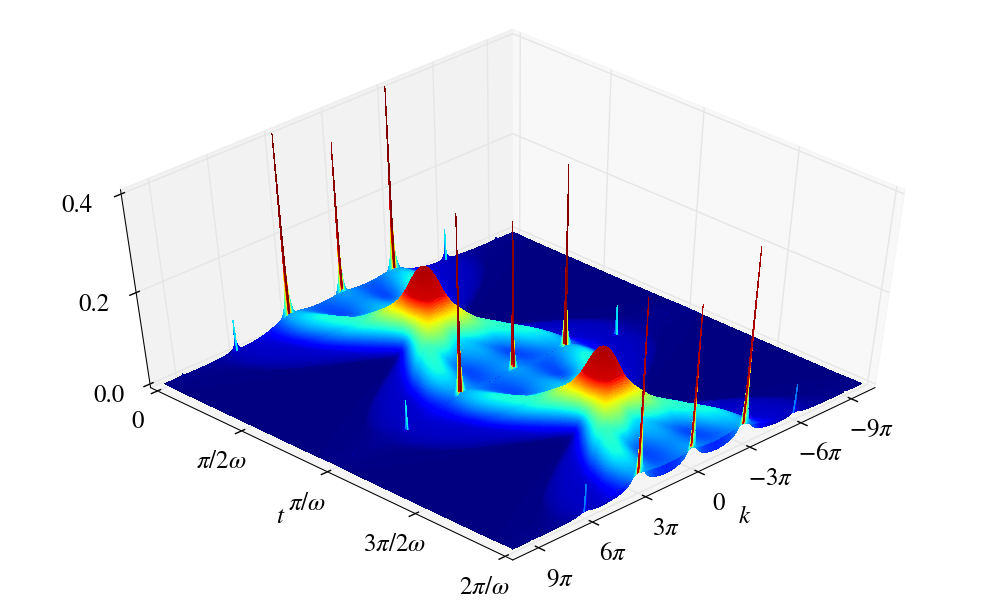}
	\end{array}$
	\end{center}
	\caption{The time evolution of the density (top) and MDF (bottom) in the trap, computed with the FB mapping for $N=50$,  $\omega = 10/N$, $A = 1.5$ and $q = 3 \pi$.}
	\label{fig:harmonic_dens_mom}
	\end{figure}

Next, we will use the FB mapping to investigate how these observations translate to the more experimentally relevant geometry of a harmonic trapping potential, with the Hamiltonian $H_{\text{trap}} = H_{\text{LL}} + \sum_{i=1}^N  \frac{1}{2} m \omega^2 x_i^2$ and $\omega$ the trapping frequency. 
The ground state SP harmonic oscillator wavefunctions are given by
\begin{align}
\psi_j(x) = \frac{1}{\sqrt{2^j j!}} \left(\frac{m \omega}{\pi}\right)^{1/4} e^{-\frac{m\omega x^2}{2}} H_j\left(\sqrt{m \omega} x\right),
\label{eq:harm_eig}
\end{align}
for $j=1,...,N$, with $H_j(x)$ denoting the Hermite polynomials. Similar to Eq.~\eqref{eq:psi_0_ring}, acting on these states with the Bragg operator $\hU_B$ leads to an additional phase of the one-body cosine potential in the SP wavefunctions. Using the propagator for the quantum harmonic oscillator (the Mehler kernel)~\cite{1866_Mehler_German}, we compute the time evolution of the SP wavefunctions \cite{SupMat}:
\begin{align}
\psi_j(x;t) =&  \sum_{\beta=-\infty}^{\infty}I_{\beta}(-iA)e^{-i\beta q \cos(\omega t)\left(x+\frac{\beta q}{2m\omega}\sin(\omega t)\right)} \notag \\
 &\;\psi_j(x+ \tfrac{\beta q}{m\omega}\sin(\omega t))  e^{-i \omega(j+\frac{1}{2})t}.
\end{align}
As a consequence of the Tonks-Girardeau limit, the SP wavefunctions are periodic in time with period $2\pi/\omega$, which is reflected in observables such as the density and the MDF. This periodicity is expected to be broken by finite-$c$ interactions and anharmonicities in the trapping potential.
The time evolution of the density and the MDF during one period is shown in Fig. \ref{fig:harmonic_dens_mom}, 
where the contributions from particles belonging to different momentum satellites are clearly distinguishable. During the initial stages of relaxation (and around multiples of $t = \tfrac{\pi}{\omega}$) the density shows strong oscillations and the initially sharply peaked MDF relaxes rapidly to a more broadened shape. This prerelaxation is well separated from the collective periodic motion due to the trap, suggesting that it is governed by the same physics as relaxation on a ring. 
%

\begin{figure}[ht]
	\includegraphics[width=0.99 \columnwidth]{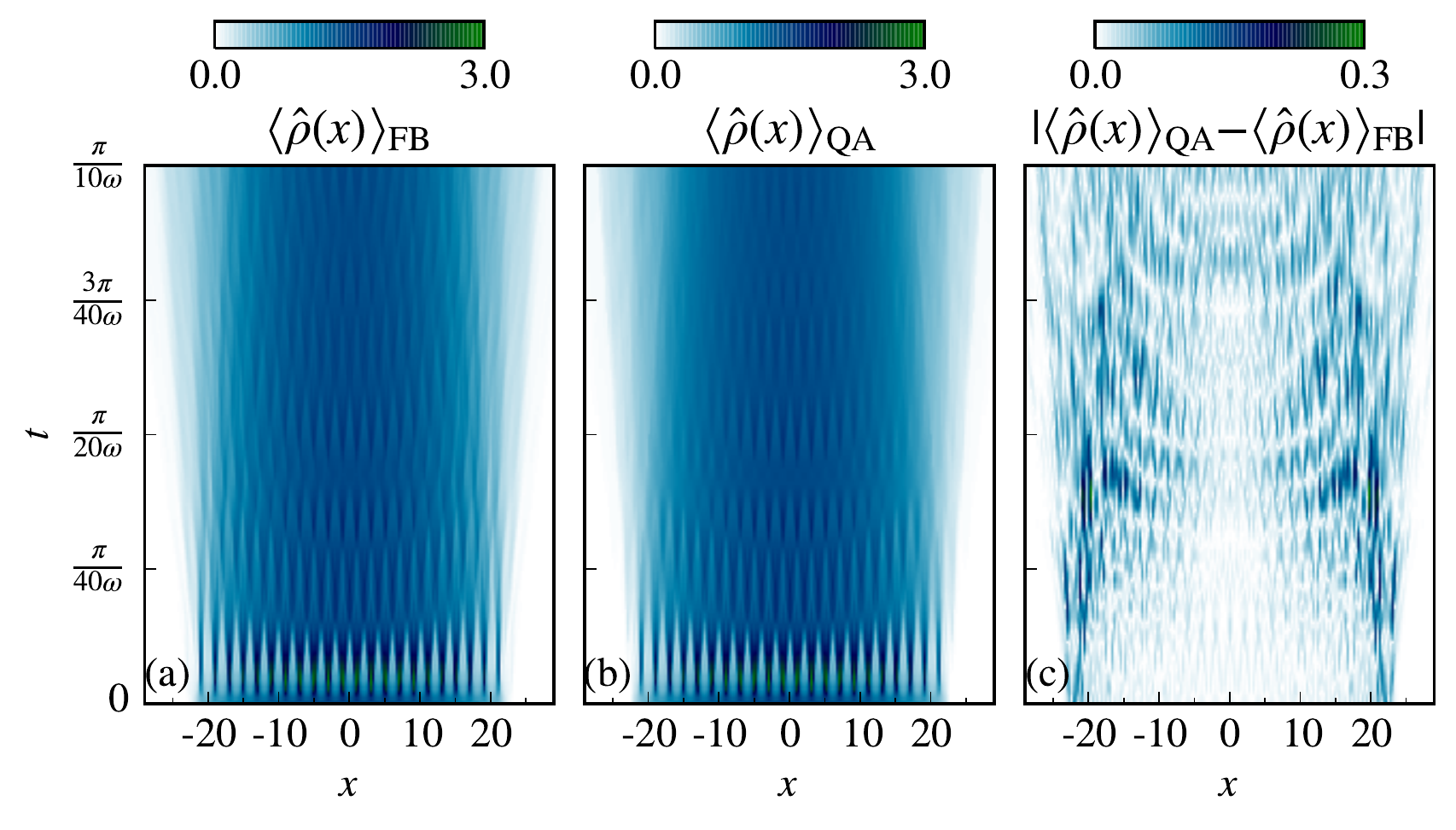}
		\caption{Time evolution of the density in a trap, computed with: the FB mapping for $N=50$ particles (left), the QA approach on a ring with an LDA accounting for the trap (middle). The difference between the two results is shown on the right. The Bragg pulse parameters are set to $A =1.5$ and $q=\pi$ with a trapping frequency $\omega  = 10/N$.}
	\label{fig: LDA_QA_Fermibose}
\end{figure}

In Fig.~\ref{fig: LDA_QA_Fermibose} the density at early stages in the oscillation cycle is compared with the density dynamics on a ring, where the latter was supplemented by a local density approximation (LDA) to account for the classical expansion of the gas in the trap ~\cite{2015_Campbell_PRL_114, SupMat}. 
The initial density profile is accurately reproduced by the LDA, except for small differences near the edges originating from the gradient in the local density not accounted for by the LDA \cite{2005_Kheruntsyan_PRA_71,2008_Amerongen_PRL_100,2011_Armijo_PRA_83,
2011_Jacqmin_PRL_106}. Note however that these differences do not stay confined to the edges and propagate towards the center as time progresses.

\begin{figure}[htp]
\includegraphics[width=0.99 \columnwidth]{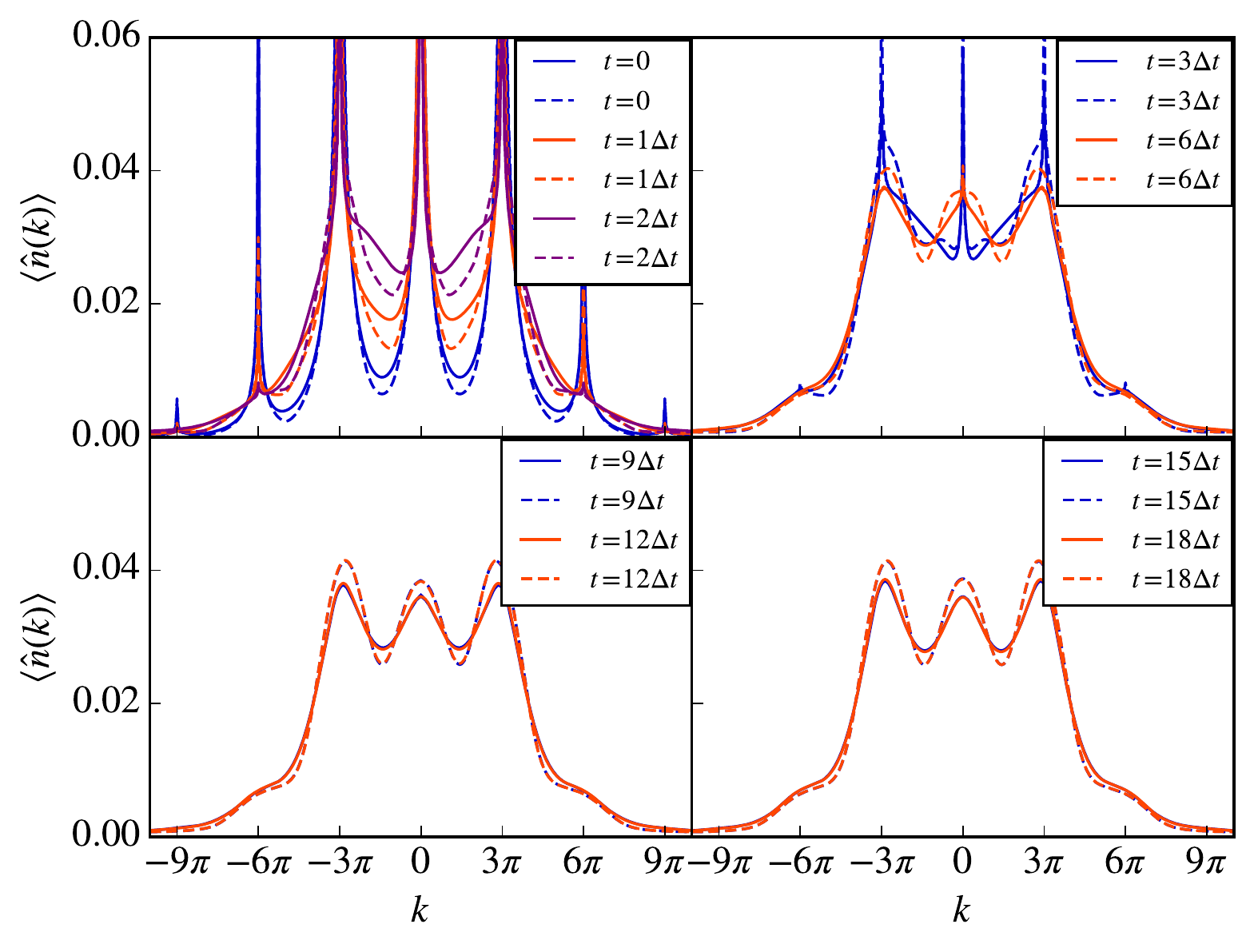}
		\caption{Time evolution of the MDF for the trap geometry (solid lines) and the ring geometry (dashed lines), obtained with the FB mapping for $N=50$ particles. The trapping frequency is set to $\omega = 10/N$, and the Bragg parameters are given by $A = 1.5$ and $q = 3\pi$. The time step $\Delta t$ is set to $\tfrac{\pi}{800\omega}$.}
	\label{fig:trap_ring_mom_cut}
\end{figure}

The short-time MDF in the trap and ring geometry is shown in Fig.~\ref{fig:trap_ring_mom_cut} up to $t = 0.0225 \pi /\omega$ . The initial distributions are nearly identical, after which the MDFs dephase in a similar fashion to a (pre)relaxed ghost-like shape. 
The strong similarities can be attributed to the short-range correlations characterizing the post-quench steady state, for which the one-body density matrix decays exponentially as a function of the distance between the particles. Large-distance effects due to the trap geometry lead to discrepancies only at low momenta $k$. 
We conclude that the short-time dynamics in a trap closely resembles the dephasing on a ring and is thus governed by the physics of hard-core interactions. The time scale associated to this (pre)relaxation is much shorter than the collective oscillations in the trap. Considering conditions similar to the Newton's cradle experiment, we estimate the short time scale to be of the order of  $10$ $\mu$s. Since this estimate is of the same order of magnitude as the pulse durations used in \cite{2006_Kinoshita_NATURE_440}, 
an interesting next step would be to extend the study of the effects of interactions to longer pulses in the Bragg regime. 

An interesting open question is how these observations extend to finite-$c$ interactions. Away from the TG limit, the initial post-pulse MDF remains a weighted sum of copies of the ground-state MDF \cite{SupMat}, leading to a decreasing spread of momenta in the satellites centered around multiples of $q$, as one goes from the hardcore limit to the BEC limit ($c\rightarrow0$). 
Although finite-$c$ dynamics is currently unattainable, we expect the time scale of the rapid relaxation to grow for smaller $c$, proportional to the inverse of the sound velocity, $v_s =  \frac{\pi n}{m} \big( 1 - \frac{4 n}{c} + \frac{12 n^2}{c^2} + \ldots \big)$.

\paragraph*{Conclusion.}
In summary, we have developed a theoretical description of the Bragg pulse for one-dimensional Bose gases and shown that the time evolution of physical observables for a Bragg pulsed Tonks-Girardeau gas in a trap is characterized by two well-separated time scales. The shortest time scale is dominated by the trap-insensitive hard-core interactions and causes a substantial broadening of the momentum distribution function well before the collective motion due to the presence of the trap sets in. Our work opens up the possibility to study the influence of interactions on more general pulse protocols and their detailed effects on experimentally relevant observables.

We thank M. Brockmann, N.J. van Druten, V. Gritsev, F. Meinert, H.-C. N\"agerl, J. Schmiedmayer, F.E. Schreck, D. Weiss and J. van Wezel for useful discussions. This work was supported by the Netherlands Organisation for Scientific Research (NWO) and the Foundation for Fundamental Research on Matter (FOM), and forms part of the activities of the Delta-Institute for Theoretical Physics (D-ITP). This research was done in part under the auspices of the CMPMS Dept. at Brookhaven National Laboratory, which in turn is supported by the U.S. Department of Energy, Office of Basic Energy Sciences, under Contract No. DE-AC02-98CH10886. We are grateful for support from the Centre de Recherches Math\'ematiques of the U. de Montr\'eal, where this work was completed.

\bibliography{literature}

\begin{thebibliography}{45}%
\makeatletter
\providecommand \@ifxundefined [1]{%
 \@ifx{#1\undefined}
}%
\providecommand \@ifnum [1]{%
 \ifnum #1\expandafter \@firstoftwo
 \else \expandafter \@secondoftwo
 \fi
}%
\providecommand \@ifx [1]{%
 \ifx #1\expandafter \@firstoftwo
 \else \expandafter \@secondoftwo
 \fi
}%
\providecommand \natexlab [1]{#1}%
\providecommand \enquote  [1]{``#1''}%
\providecommand \bibnamefont  [1]{#1}%
\providecommand \bibfnamefont [1]{#1}%
\providecommand \citenamefont [1]{#1}%
\providecommand \href@noop [0]{\@secondoftwo}%
\providecommand \href [0]{\begingroup \@sanitize@url \@href}%
\providecommand \@href[1]{\@@startlink{#1}\@@href}%
\providecommand \@@href[1]{\endgroup#1\@@endlink}%
\providecommand \@sanitize@url [0]{\catcode `\\12\catcode `\$12\catcode
  `\&12\catcode `\#12\catcode `\^12\catcode `\_12\catcode `\%12\relax}%
\providecommand \@@startlink[1]{}%
\providecommand \@@endlink[0]{}%
\providecommand \url  [0]{\begingroup\@sanitize@url \@url }%
\providecommand \@url [1]{\endgroup\@href {#1}{\urlprefix }}%
\providecommand \urlprefix  [0]{URL }%
\providecommand \Eprint [0]{\href }%
\providecommand \doibase [0]{http://dx.doi.org/}%
\providecommand \selectlanguage [0]{\@gobble}%
\providecommand \bibinfo  [0]{\@secondoftwo}%
\providecommand \bibfield  [0]{\@secondoftwo}%
\providecommand \translation [1]{[#1]}%
\providecommand \BibitemOpen [0]{}%
\providecommand \bibitemStop [0]{}%
\providecommand \bibitemNoStop [0]{.\EOS\space}%
\providecommand \EOS [0]{\spacefactor3000\relax}%
\providecommand \BibitemShut  [1]{\csname bibitem#1\endcsname}%
\let\auto@bib@innerbib\@empty
\bibitem [{\citenamefont {Bloch}\ \emph {et~al.}(2008)\citenamefont {Bloch},
  \citenamefont {Dalibard},\ and\ \citenamefont {Zwerger}}]{2008_Bloch_RMP_80}%
  \BibitemOpen
  \bibfield  {author} {\bibinfo {author} {\bibfnamefont {I.}~\bibnamefont
  {Bloch}}, \bibinfo {author} {\bibfnamefont {J.}~\bibnamefont {Dalibard}}, \
  and\ \bibinfo {author} {\bibfnamefont {W.}~\bibnamefont {Zwerger}},\ }\href
  {\doibase 10.1103/RevModPhys.80.885} {\bibfield  {journal} {\bibinfo
  {journal} {Rev. Mod. Phys.}\ }\textbf {\bibinfo {volume} {80}},\ \bibinfo
  {pages} {885} (\bibinfo {year} {2008})}\BibitemShut {NoStop}%
\bibitem [{\citenamefont {Cazalilla}\ \emph {et~al.}(2011)\citenamefont
  {Cazalilla}, \citenamefont {Citro}, \citenamefont {Giamarchi}, \citenamefont
  {Orignac},\ and\ \citenamefont {Rigol}}]{2011_Cazalilla_RMP_83}%
  \BibitemOpen
  \bibfield  {author} {\bibinfo {author} {\bibfnamefont {M.~A.}\ \bibnamefont
  {Cazalilla}}, \bibinfo {author} {\bibfnamefont {R.}~\bibnamefont {Citro}},
  \bibinfo {author} {\bibfnamefont {T.}~\bibnamefont {Giamarchi}}, \bibinfo
  {author} {\bibfnamefont {E.}~\bibnamefont {Orignac}}, \ and\ \bibinfo
  {author} {\bibfnamefont {M.}~\bibnamefont {Rigol}},\ }\href {\doibase
  10.1103/RevModPhys.83.1405} {\bibfield  {journal} {\bibinfo  {journal} {Rev.
  Mod. Phys.}\ }\textbf {\bibinfo {volume} {83}},\ \bibinfo {pages} {1405}
  (\bibinfo {year} {2011})}\BibitemShut {NoStop}%
\bibitem [{\citenamefont {Giamarchi}(2004)}]{GiamarchiBOOK}%
  \BibitemOpen
  \bibfield  {author} {\bibinfo {author} {\bibfnamefont {T.}~\bibnamefont
  {Giamarchi}},\ }\href@noop {} {\emph {\bibinfo {title} {Quantum Physics in
  One Dimension}}}\ (\bibinfo  {publisher} {Oxford University Press},\ \bibinfo
  {year} {2004})\BibitemShut {NoStop}%
\bibitem [{\citenamefont {Martin}\ \emph {et~al.}(1988)\citenamefont {Martin},
  \citenamefont {Oldaker}, \citenamefont {Miklich},\ and\ \citenamefont
  {Pritchard}}]{1988_Martin_PRL_60}%
  \BibitemOpen
  \bibfield  {author} {\bibinfo {author} {\bibfnamefont {P.~J.}\ \bibnamefont
  {Martin}}, \bibinfo {author} {\bibfnamefont {B.~G.}\ \bibnamefont {Oldaker}},
  \bibinfo {author} {\bibfnamefont {A.~H.}\ \bibnamefont {Miklich}}, \ and\
  \bibinfo {author} {\bibfnamefont {D.~E.}\ \bibnamefont {Pritchard}},\ }\href
  {\doibase 10.1103/PhysRevLett.60.515} {\bibfield  {journal} {\bibinfo
  {journal} {Phys. Rev. Lett.}\ }\textbf {\bibinfo {volume} {60}},\ \bibinfo
  {pages} {515} (\bibinfo {year} {1988})}\BibitemShut {NoStop}%
\bibitem [{\citenamefont {Stenger}\ \emph {et~al.}(1999)\citenamefont
  {Stenger}, \citenamefont {Inouye}, \citenamefont {Chikkatur}, \citenamefont
  {Stamper-Kurn}, \citenamefont {Pritchard},\ and\ \citenamefont
  {Ketterle}}]{1999_Stenger_PRL_82}%
  \BibitemOpen
  \bibfield  {author} {\bibinfo {author} {\bibfnamefont {J.}~\bibnamefont
  {Stenger}}, \bibinfo {author} {\bibfnamefont {S.}~\bibnamefont {Inouye}},
  \bibinfo {author} {\bibfnamefont {A.~P.}\ \bibnamefont {Chikkatur}}, \bibinfo
  {author} {\bibfnamefont {D.~M.}\ \bibnamefont {Stamper-Kurn}}, \bibinfo
  {author} {\bibfnamefont {D.~E.}\ \bibnamefont {Pritchard}}, \ and\ \bibinfo
  {author} {\bibfnamefont {W.}~\bibnamefont {Ketterle}},\ }\href {\doibase
  10.1103/PhysRevLett.82.4569} {\bibfield  {journal} {\bibinfo  {journal}
  {Phys. Rev. Lett.}\ }\textbf {\bibinfo {volume} {82}},\ \bibinfo {pages}
  {4569} (\bibinfo {year} {1999})}\BibitemShut {NoStop}%
\bibitem [{\citenamefont {Ozeri}\ \emph {et~al.}(2005)\citenamefont {Ozeri},
  \citenamefont {Katz}, \citenamefont {Steinhauer},\ and\ \citenamefont
  {Davidson}}]{2005_Ozeri_RMP_77}%
  \BibitemOpen
  \bibfield  {author} {\bibinfo {author} {\bibfnamefont {R.}~\bibnamefont
  {Ozeri}}, \bibinfo {author} {\bibfnamefont {N.}~\bibnamefont {Katz}},
  \bibinfo {author} {\bibfnamefont {J.}~\bibnamefont {Steinhauer}}, \ and\
  \bibinfo {author} {\bibfnamefont {N.}~\bibnamefont {Davidson}},\ }\href
  {\doibase 10.1103/RevModPhys.77.187} {\bibfield  {journal} {\bibinfo
  {journal} {Rev. Mod. Phys.}\ }\textbf {\bibinfo {volume} {77}},\ \bibinfo
  {pages} {187} (\bibinfo {year} {2005})}\BibitemShut {NoStop}%
\bibitem [{\citenamefont {Wang}\ \emph {et~al.}(2005)\citenamefont {Wang},
  \citenamefont {Anderson}, \citenamefont {Bright}, \citenamefont {Cornell},
  \citenamefont {Diot}, \citenamefont {Kishimoto}, \citenamefont {Prentiss},
  \citenamefont {Saravanan}, \citenamefont {Segal},\ and\ \citenamefont
  {Wu}}]{2005_Wang_PRL_94}%
  \BibitemOpen
  \bibfield  {author} {\bibinfo {author} {\bibfnamefont {Y.-J.}\ \bibnamefont
  {Wang}}, \bibinfo {author} {\bibfnamefont {D.~Z.}\ \bibnamefont {Anderson}},
  \bibinfo {author} {\bibfnamefont {V.~M.}\ \bibnamefont {Bright}}, \bibinfo
  {author} {\bibfnamefont {E.~A.}\ \bibnamefont {Cornell}}, \bibinfo {author}
  {\bibfnamefont {Q.}~\bibnamefont {Diot}}, \bibinfo {author} {\bibfnamefont
  {T.}~\bibnamefont {Kishimoto}}, \bibinfo {author} {\bibfnamefont
  {M.}~\bibnamefont {Prentiss}}, \bibinfo {author} {\bibfnamefont {R.~A.}\
  \bibnamefont {Saravanan}}, \bibinfo {author} {\bibfnamefont {S.~R.}\
  \bibnamefont {Segal}}, \ and\ \bibinfo {author} {\bibfnamefont
  {S.}~\bibnamefont {Wu}},\ }\href {\doibase 10.1103/PhysRevLett.94.090405}
  {\bibfield  {journal} {\bibinfo  {journal} {Phys. Rev. Lett.}\ }\textbf
  {\bibinfo {volume} {94}},\ \bibinfo {pages} {090405} (\bibinfo {year}
  {2005})}\BibitemShut {NoStop}%
\bibitem [{\citenamefont {Wu}\ \emph {et~al.}(2005)\citenamefont {Wu},
  \citenamefont {Wang}, \citenamefont {Diot},\ and\ \citenamefont
  {Prentiss}}]{2005_Wu_PRA_71}%
  \BibitemOpen
  \bibfield  {author} {\bibinfo {author} {\bibfnamefont {S.}~\bibnamefont
  {Wu}}, \bibinfo {author} {\bibfnamefont {Y.-J.}\ \bibnamefont {Wang}},
  \bibinfo {author} {\bibfnamefont {Q.}~\bibnamefont {Diot}}, \ and\ \bibinfo
  {author} {\bibfnamefont {M.}~\bibnamefont {Prentiss}},\ }\href {\doibase
  10.1103/PhysRevA.71.043602} {\bibfield  {journal} {\bibinfo  {journal} {Phys.
  Rev. A}\ }\textbf {\bibinfo {volume} {71}},\ \bibinfo {pages} {043602}
  (\bibinfo {year} {2005})}\BibitemShut {NoStop}%
\bibitem [{\citenamefont {Lieb}\ and\ \citenamefont
  {Liniger}(1963)}]{1963_Lieb_PR_130}%
  \BibitemOpen
  \bibfield  {author} {\bibinfo {author} {\bibfnamefont {E.~H.}\ \bibnamefont
  {Lieb}}\ and\ \bibinfo {author} {\bibfnamefont {W.}~\bibnamefont {Liniger}},\
  }\href {\doibase 10.1103/PhysRev.130.1605} {\bibfield  {journal} {\bibinfo
  {journal} {Phys. Rev.}\ }\textbf {\bibinfo {volume} {130}},\ \bibinfo {pages}
  {1605} (\bibinfo {year} {1963})}\BibitemShut {NoStop}%
\bibitem [{\citenamefont {Olshanii}(1998)}]{1998_Olshanii_PRL_81}%
  \BibitemOpen
  \bibfield  {author} {\bibinfo {author} {\bibfnamefont {M.}~\bibnamefont
  {Olshanii}},\ }\href {\doibase 10.1103/PhysRevLett.81.938} {\bibfield
  {journal} {\bibinfo  {journal} {Phys. Rev. Lett.}\ }\textbf {\bibinfo
  {volume} {81}},\ \bibinfo {pages} {938} (\bibinfo {year} {1998})}\BibitemShut
  {NoStop}%
\bibitem [{\citenamefont {Kinoshita}\ \emph {et~al.}(2006)\citenamefont
  {Kinoshita}, \citenamefont {Wenger},\ and\ \citenamefont
  {Weiss}}]{2006_Kinoshita_NATURE_440}%
  \BibitemOpen
  \bibfield  {author} {\bibinfo {author} {\bibfnamefont {T.}~\bibnamefont
  {Kinoshita}}, \bibinfo {author} {\bibfnamefont {T.}~\bibnamefont {Wenger}}, \
  and\ \bibinfo {author} {\bibfnamefont {D.~S.}\ \bibnamefont {Weiss}},\ }\href
  {\doibase 10.1038/nature04693} {\bibfield  {journal} {\bibinfo  {journal}
  {Nature}\ }\textbf {\bibinfo {volume} {440}},\ \bibinfo {pages} {900}
  (\bibinfo {year} {2006})}\BibitemShut {NoStop}%
\bibitem [{\citenamefont {Fabbri}\ \emph {et~al.}(2015)\citenamefont {Fabbri},
  \citenamefont {Panfil}, \citenamefont {Cl\'ement}, \citenamefont {Fallani},
  \citenamefont {Inguscio}, \citenamefont {Fort},\ and\ \citenamefont
  {Caux}}]{2015_Fabbri_PRA_91}%
  \BibitemOpen
  \bibfield  {author} {\bibinfo {author} {\bibfnamefont {N.}~\bibnamefont
  {Fabbri}}, \bibinfo {author} {\bibfnamefont {M.}~\bibnamefont {Panfil}},
  \bibinfo {author} {\bibfnamefont {D.}~\bibnamefont {Cl\'ement}}, \bibinfo
  {author} {\bibfnamefont {L.}~\bibnamefont {Fallani}}, \bibinfo {author}
  {\bibfnamefont {M.}~\bibnamefont {Inguscio}}, \bibinfo {author}
  {\bibfnamefont {C.}~\bibnamefont {Fort}}, \ and\ \bibinfo {author}
  {\bibfnamefont {J.-S.}\ \bibnamefont {Caux}},\ }\href {\doibase
  10.1103/PhysRevA.91.043617} {\bibfield  {journal} {\bibinfo  {journal} {Phys.
  Rev. A}\ }\textbf {\bibinfo {volume} {91}},\ \bibinfo {pages} {043617}
  (\bibinfo {year} {2015})}\BibitemShut {NoStop}%
\bibitem [{\citenamefont {Meinert}\ \emph {et~al.}(2015)\citenamefont
  {Meinert}, \citenamefont {Panfil}, \citenamefont {Mark}, \citenamefont
  {Lauber}, \citenamefont {Caux},\ and\ \citenamefont
  {N\"agerl}}]{2015_Meinert_PRL_115}%
  \BibitemOpen
  \bibfield  {author} {\bibinfo {author} {\bibfnamefont {F.}~\bibnamefont
  {Meinert}}, \bibinfo {author} {\bibfnamefont {M.}~\bibnamefont {Panfil}},
  \bibinfo {author} {\bibfnamefont {M.~J.}\ \bibnamefont {Mark}}, \bibinfo
  {author} {\bibfnamefont {K.}~\bibnamefont {Lauber}}, \bibinfo {author}
  {\bibfnamefont {J.-S.}\ \bibnamefont {Caux}}, \ and\ \bibinfo {author}
  {\bibfnamefont {H.-C.}\ \bibnamefont {N\"agerl}},\ }\href {\doibase
  10.1103/PhysRevLett.115.085301} {\bibfield  {journal} {\bibinfo  {journal}
  {Phys. Rev. Lett.}\ }\textbf {\bibinfo {volume} {115}},\ \bibinfo {pages}
  {085301} (\bibinfo {year} {2015})}\BibitemShut {NoStop}%
\bibitem [{\citenamefont {Panfil}\ and\ \citenamefont
  {Caux}(2014)}]{2014_Panfil_PRA_89}%
  \BibitemOpen
  \bibfield  {author} {\bibinfo {author} {\bibfnamefont {M.}~\bibnamefont
  {Panfil}}\ and\ \bibinfo {author} {\bibfnamefont {J.-S.}\ \bibnamefont
  {Caux}},\ }\href {\doibase 10.1103/PhysRevA.89.033605} {\bibfield  {journal}
  {\bibinfo  {journal} {Phys. Rev. A}\ }\textbf {\bibinfo {volume} {89}},\
  \bibinfo {pages} {033605} (\bibinfo {year} {2014})}\BibitemShut {NoStop}%
\bibitem [{\citenamefont {Tonks}(1936)}]{1936_Tonks_PR_50}%
  \BibitemOpen
  \bibfield  {author} {\bibinfo {author} {\bibfnamefont {L.}~\bibnamefont
  {Tonks}},\ }\href {\doibase 10.1103/PhysRev.50.955} {\bibfield  {journal}
  {\bibinfo  {journal} {Phys. Rev.}\ }\textbf {\bibinfo {volume} {50}},\
  \bibinfo {pages} {955} (\bibinfo {year} {1936})}\BibitemShut {NoStop}%
\bibitem [{\citenamefont {Girardeau}(1960)}]{1960_Girardeau_JMP_1}%
  \BibitemOpen
  \bibfield  {author} {\bibinfo {author} {\bibfnamefont {M.}~\bibnamefont
  {Girardeau}},\ }\href {\doibase 10.1063/1.1703687} {\bibfield  {journal}
  {\bibinfo  {journal} {J. Math. Phys.}\ }\textbf {\bibinfo {volume} {1}},\
  \bibinfo {pages} {516} (\bibinfo {year} {1960})}\BibitemShut {NoStop}%
\bibitem [{\citenamefont {Haller}\ \emph {et~al.}(2009)\citenamefont {Haller},
  \citenamefont {Gustavsson}, \citenamefont {Mark}, \citenamefont {Danzl},
  \citenamefont {Hart}, \citenamefont {Pupillo},\ and\ \citenamefont
  {N{\"a}gerl}}]{2009_Haller_SCIENCE_325}%
  \BibitemOpen
  \bibfield  {author} {\bibinfo {author} {\bibfnamefont {E.}~\bibnamefont
  {Haller}}, \bibinfo {author} {\bibfnamefont {M.}~\bibnamefont {Gustavsson}},
  \bibinfo {author} {\bibfnamefont {M.~J.}\ \bibnamefont {Mark}}, \bibinfo
  {author} {\bibfnamefont {J.~G.}\ \bibnamefont {Danzl}}, \bibinfo {author}
  {\bibfnamefont {R.}~\bibnamefont {Hart}}, \bibinfo {author} {\bibfnamefont
  {G.}~\bibnamefont {Pupillo}}, \ and\ \bibinfo {author} {\bibfnamefont
  {H.-C.}\ \bibnamefont {N{\"a}gerl}},\ }\href {\doibase
  10.1126/science.1175850} {\bibfield  {journal} {\bibinfo  {journal}
  {Science}\ }\textbf {\bibinfo {volume} {325}},\ \bibinfo {pages} {1224}
  (\bibinfo {year} {2009})}\BibitemShut {NoStop}%
\bibitem [{\citenamefont {Minguzzi}\ and\ \citenamefont
  {Gangardt}(2005)}]{2005_Minguzzi_PRL_94}%
  \BibitemOpen
  \bibfield  {author} {\bibinfo {author} {\bibfnamefont {A.}~\bibnamefont
  {Minguzzi}}\ and\ \bibinfo {author} {\bibfnamefont {D.~M.}\ \bibnamefont
  {Gangardt}},\ }\href {\doibase 10.1103/PhysRevLett.94.240404} {\bibfield
  {journal} {\bibinfo  {journal} {Phys. Rev. Lett.}\ }\textbf {\bibinfo
  {volume} {94}},\ \bibinfo {pages} {240404} (\bibinfo {year}
  {2005})}\BibitemShut {NoStop}%
\bibitem [{\citenamefont {Pezer}\ and\ \citenamefont
  {Buljan}(2007)}]{2007_Pezer_PRL_98}%
  \BibitemOpen
  \bibfield  {author} {\bibinfo {author} {\bibfnamefont {R.}~\bibnamefont
  {Pezer}}\ and\ \bibinfo {author} {\bibfnamefont {H.}~\bibnamefont {Buljan}},\
  }\href {\doibase 10.1103/PhysRevLett.98.240403} {\bibfield  {journal}
  {\bibinfo  {journal} {Phys. Rev. Lett.}\ }\textbf {\bibinfo {volume} {98}},\
  \bibinfo {pages} {240403} (\bibinfo {year} {2007})}\BibitemShut {NoStop}%
\bibitem [{\citenamefont {Gangardt}\ and\ \citenamefont
  {Pustilnik}(2008)}]{2008_Gangardt_PRA_77}%
  \BibitemOpen
  \bibfield  {author} {\bibinfo {author} {\bibfnamefont {D.~M.}\ \bibnamefont
  {Gangardt}}\ and\ \bibinfo {author} {\bibfnamefont {M.}~\bibnamefont
  {Pustilnik}},\ }\href {\doibase 10.1103/PhysRevA.77.041604} {\bibfield
  {journal} {\bibinfo  {journal} {Phys. Rev. A}\ }\textbf {\bibinfo {volume}
  {77}},\ \bibinfo {pages} {041604} (\bibinfo {year} {2008})}\BibitemShut
  {NoStop}%
\bibitem [{\citenamefont {Muth}\ \emph {et~al.}(2010)\citenamefont {Muth},
  \citenamefont {Schmidt},\ and\ \citenamefont
  {Fleischhauer}}]{2010_Muth_NJP_12}%
  \BibitemOpen
  \bibfield  {author} {\bibinfo {author} {\bibfnamefont {D.}~\bibnamefont
  {Muth}}, \bibinfo {author} {\bibfnamefont {B.}~\bibnamefont {Schmidt}}, \
  and\ \bibinfo {author} {\bibfnamefont {M.}~\bibnamefont {Fleischhauer}},\
  }\href {http://stacks.iop.org/1367-2630/12/i=8/a=083065} {\bibfield
  {journal} {\bibinfo  {journal} {New Journal of Physics}\ }\textbf {\bibinfo
  {volume} {12}},\ \bibinfo {pages} {083065} (\bibinfo {year}
  {2010})}\BibitemShut {NoStop}%
\bibitem [{\citenamefont {Schenke}\ \emph {et~al.}(2011)\citenamefont
  {Schenke}, \citenamefont {Minguzzi},\ and\ \citenamefont
  {Hekking}}]{2011_Schenke_PRA_84}%
  \BibitemOpen
  \bibfield  {author} {\bibinfo {author} {\bibfnamefont {C.}~\bibnamefont
  {Schenke}}, \bibinfo {author} {\bibfnamefont {A.}~\bibnamefont {Minguzzi}}, \
  and\ \bibinfo {author} {\bibfnamefont {F.~W.~J.}\ \bibnamefont {Hekking}},\
  }\href {\doibase 10.1103/PhysRevA.84.053636} {\bibfield  {journal} {\bibinfo
  {journal} {Phys. Rev. A}\ }\textbf {\bibinfo {volume} {84}},\ \bibinfo
  {pages} {053636} (\bibinfo {year} {2011})}\BibitemShut {NoStop}%
\bibitem [{\citenamefont {Astrakharchik}\ and\ \citenamefont
  {Pitaevskii}(2013)}]{2013_Astrakharchik_EPL_102}%
  \BibitemOpen
  \bibfield  {author} {\bibinfo {author} {\bibfnamefont {G.~E.}\ \bibnamefont
  {Astrakharchik}}\ and\ \bibinfo {author} {\bibfnamefont {L.~P.}\ \bibnamefont
  {Pitaevskii}},\ }\href {http://stacks.iop.org/0295-5075/102/i=3/a=30004}
  {\bibfield  {journal} {\bibinfo  {journal} {Europhys. Lett.}\ }\textbf
  {\bibinfo {volume} {102}},\ \bibinfo {pages} {30004} (\bibinfo {year}
  {2013})}\BibitemShut {NoStop}%
\bibitem [{\citenamefont {Collura}\ \emph {et~al.}(2013)\citenamefont
  {Collura}, \citenamefont {Sotiriadis},\ and\ \citenamefont
  {Calabrese}}]{2013_Collura_PRL_110}%
  \BibitemOpen
  \bibfield  {author} {\bibinfo {author} {\bibfnamefont {M.}~\bibnamefont
  {Collura}}, \bibinfo {author} {\bibfnamefont {S.}~\bibnamefont {Sotiriadis}},
  \ and\ \bibinfo {author} {\bibfnamefont {P.}~\bibnamefont {Calabrese}},\
  }\href {\doibase 10.1103/PhysRevLett.110.245301} {\bibfield  {journal}
  {\bibinfo  {journal} {Phys. Rev. Lett.}\ }\textbf {\bibinfo {volume} {110}},\
  \bibinfo {pages} {245301} (\bibinfo {year} {2013})}\BibitemShut {NoStop}%
\bibitem [{\citenamefont {Quinn}\ and\ \citenamefont
  {Haque}(2014)}]{2014_Quinn_PRA_90}%
  \BibitemOpen
  \bibfield  {author} {\bibinfo {author} {\bibfnamefont {E.}~\bibnamefont
  {Quinn}}\ and\ \bibinfo {author} {\bibfnamefont {M.}~\bibnamefont {Haque}},\
  }\href {\doibase 10.1103/PhysRevA.90.053609} {\bibfield  {journal} {\bibinfo
  {journal} {Phys. Rev. A}\ }\textbf {\bibinfo {volume} {90}},\ \bibinfo
  {pages} {053609} (\bibinfo {year} {2014})}\BibitemShut {NoStop}%
\bibitem [{\citenamefont {{Cartarius}}\ \emph {et~al.}()\citenamefont
  {{Cartarius}}, \citenamefont {{Kawasaki}},\ and\ \citenamefont
  {{Minguzzi}}}]{2015_Cartarius_arXiv}%
  \BibitemOpen
  \bibfield  {author} {\bibinfo {author} {\bibfnamefont {F.}~\bibnamefont
  {{Cartarius}}}, \bibinfo {author} {\bibfnamefont {E.}~\bibnamefont
  {{Kawasaki}}}, \ and\ \bibinfo {author} {\bibfnamefont {A.}~\bibnamefont
  {{Minguzzi}}},\ }\href@noop {} {\ }\Eprint {http://arxiv.org/abs/1505.01009}
  {arXiv:1505.01009} \BibitemShut {NoStop}%
\bibitem [{\citenamefont {{Kapitza}}\ and\ \citenamefont
  {{Dirac}}(1933)}]{1933_Kapitza_Proc}%
  \BibitemOpen
  \bibfield  {author} {\bibinfo {author} {\bibfnamefont {P.~L.}\ \bibnamefont
  {{Kapitza}}}\ and\ \bibinfo {author} {\bibfnamefont {P.~A.~M.}\ \bibnamefont
  {{Dirac}}},\ }\href {\doibase 10.1017/S0305004100011105} {\bibfield
  {journal} {\bibinfo  {journal} {Proceedings of the Cambridge Philosophical
  Society}\ }\textbf {\bibinfo {volume} {29}},\ \bibinfo {pages} {297}
  (\bibinfo {year} {1933})}\BibitemShut {NoStop}%
\bibitem [{\citenamefont {Freimund}\ \emph {et~al.}(2001)\citenamefont
  {Freimund}, \citenamefont {Aflatooni},\ and\ \citenamefont
  {Batelaan}}]{2001_Freimund_Nat}%
  \BibitemOpen
  \bibfield  {author} {\bibinfo {author} {\bibfnamefont {D.~L.}\ \bibnamefont
  {Freimund}}, \bibinfo {author} {\bibfnamefont {K.}~\bibnamefont {Aflatooni}},
  \ and\ \bibinfo {author} {\bibfnamefont {H.}~\bibnamefont {Batelaan}},\
  }\href {\doibase 10.1038/35093065} {\bibfield  {journal} {\bibinfo  {journal}
  {Nature}\ }\textbf {\bibinfo {volume} {413}},\ \bibinfo {pages} {142}
  (\bibinfo {year} {2001})}\BibitemShut {NoStop}%
\bibitem [{\citenamefont {Calabrese}\ and\ \citenamefont
  {Cardy}(2006)}]{2006_Calabrese_PRL_96}%
  \BibitemOpen
  \bibfield  {author} {\bibinfo {author} {\bibfnamefont {P.}~\bibnamefont
  {Calabrese}}\ and\ \bibinfo {author} {\bibfnamefont {J.}~\bibnamefont
  {Cardy}},\ }\href@noop {} {\bibfield  {journal} {\bibinfo  {journal} {Phys.
  Rev. Lett.}\ }\textbf {\bibinfo {volume} {96}},\ \bibinfo {pages} {136801}
  (\bibinfo {year} {2006})}\BibitemShut {NoStop}%
\bibitem [{\citenamefont {Rigol}\ \emph {et~al.}(2007)\citenamefont {Rigol},
  \citenamefont {Dunjko}, \citenamefont {Yurovsky},\ and\ \citenamefont
  {Olshanii}}]{2007_Rigol_PRL_98}%
  \BibitemOpen
  \bibfield  {author} {\bibinfo {author} {\bibfnamefont {M.}~\bibnamefont
  {Rigol}}, \bibinfo {author} {\bibfnamefont {V.}~\bibnamefont {Dunjko}},
  \bibinfo {author} {\bibfnamefont {V.}~\bibnamefont {Yurovsky}}, \ and\
  \bibinfo {author} {\bibfnamefont {M.}~\bibnamefont {Olshanii}},\ }\href@noop
  {} {\bibfield  {journal} {\bibinfo  {journal} {Phys. Rev. Lett.}\ }\textbf
  {\bibinfo {volume} {98}},\ \bibinfo {pages} {050405} (\bibinfo {year}
  {2007})}\BibitemShut {NoStop}%
\bibitem [{\citenamefont {Polkovnikov}\ \emph {et~al.}(2011)\citenamefont
  {Polkovnikov}, \citenamefont {Sengupta}, \citenamefont {Silva},\ and\
  \citenamefont {Vengalattore}}]{2011_Polkovnikov_RMP_83}%
  \BibitemOpen
  \bibfield  {author} {\bibinfo {author} {\bibfnamefont {A.}~\bibnamefont
  {Polkovnikov}}, \bibinfo {author} {\bibfnamefont {K.}~\bibnamefont
  {Sengupta}}, \bibinfo {author} {\bibfnamefont {A.}~\bibnamefont {Silva}}, \
  and\ \bibinfo {author} {\bibfnamefont {M.}~\bibnamefont {Vengalattore}},\
  }\href@noop {} {\bibfield  {journal} {\bibinfo  {journal} {Rev. Mod. Phys.}\
  }\textbf {\bibinfo {volume} {83}},\ \bibinfo {pages} {863} (\bibinfo {year}
  {2011})}\BibitemShut {NoStop}%
\bibitem [{\citenamefont {Sapiro}\ \emph {et~al.}(2009)\citenamefont {Sapiro},
  \citenamefont {Zhang},\ and\ \citenamefont {Raithel}}]{2009_Sapiro_PRA_79}%
  \BibitemOpen
  \bibfield  {author} {\bibinfo {author} {\bibfnamefont {R.~E.}\ \bibnamefont
  {Sapiro}}, \bibinfo {author} {\bibfnamefont {R.}~\bibnamefont {Zhang}}, \
  and\ \bibinfo {author} {\bibfnamefont {G.}~\bibnamefont {Raithel}},\ }\href
  {\doibase 10.1103/PhysRevA.79.043630} {\bibfield  {journal} {\bibinfo
  {journal} {Phys. Rev. A}\ }\textbf {\bibinfo {volume} {79}},\ \bibinfo
  {pages} {043630} (\bibinfo {year} {2009})}\BibitemShut {NoStop}%
\bibitem [{\citenamefont {Rigol}\ \emph {et~al.}(2008)\citenamefont {Rigol},
  \citenamefont {Dunjko},\ and\ \citenamefont
  {Olshanii}}]{2008_Rigol_NATURE_452}%
  \BibitemOpen
  \bibfield  {author} {\bibinfo {author} {\bibfnamefont {M.}~\bibnamefont
  {Rigol}}, \bibinfo {author} {\bibfnamefont {V.}~\bibnamefont {Dunjko}}, \
  and\ \bibinfo {author} {\bibfnamefont {M.}~\bibnamefont {Olshanii}},\
  }\href@noop {} {\bibfield  {journal} {\bibinfo  {journal} {Nature}\ }\textbf
  {\bibinfo {volume} {452}},\ \bibinfo {pages} {854} (\bibinfo {year}
  {2008})}\BibitemShut {NoStop}%
\bibitem [{\citenamefont {Caux}\ and\ \citenamefont
  {Essler}(2013)}]{2013_Caux_PRL_110}%
  \BibitemOpen
  \bibfield  {author} {\bibinfo {author} {\bibfnamefont {J.-S.}\ \bibnamefont
  {Caux}}\ and\ \bibinfo {author} {\bibfnamefont {F.~H.~L.}\ \bibnamefont
  {Essler}},\ }\href@noop {} {\bibfield  {journal} {\bibinfo  {journal} {Phys.
  Rev. Lett.}\ }\textbf {\bibinfo {volume} {110}},\ \bibinfo {pages} {257203}
  (\bibinfo {year} {2013})}\BibitemShut {NoStop}%
\bibitem [{\citenamefont {De~Nardis}\ \emph {et~al.}(2014)\citenamefont
  {De~Nardis}, \citenamefont {Wouters}, \citenamefont {Brockmann},\ and\
  \citenamefont {Caux}}]{2014_DeNardis_PRA_89}%
  \BibitemOpen
  \bibfield  {author} {\bibinfo {author} {\bibfnamefont {J.}~\bibnamefont
  {De~Nardis}}, \bibinfo {author} {\bibfnamefont {B.}~\bibnamefont {Wouters}},
  \bibinfo {author} {\bibfnamefont {M.}~\bibnamefont {Brockmann}}, \ and\
  \bibinfo {author} {\bibfnamefont {J.-S.}\ \bibnamefont {Caux}},\ }\href@noop
  {} {\bibfield  {journal} {\bibinfo  {journal} {Phys. Rev. A}\ }\textbf
  {\bibinfo {volume} {89}},\ \bibinfo {pages} {033601} (\bibinfo {year}
  {2014})}\BibitemShut {NoStop}%
\bibitem [{Sup()}]{SupMat}%
  \BibitemOpen
  \href@noop {} {}\bibinfo {note} {See Supplemental Material}\BibitemShut
  {NoStop}%
\bibitem [{\citenamefont {Bevilacqua}\ \emph {et~al.}(2011)\citenamefont
  {Bevilacqua}, \citenamefont {Biancalana}, \citenamefont {Dancheva},
  \citenamefont {Mansour},\ and\ \citenamefont {Moi}}]{2011_Bevilacqua_JMP_52}%
  \BibitemOpen
  \bibfield  {author} {\bibinfo {author} {\bibfnamefont {G.}~\bibnamefont
  {Bevilacqua}}, \bibinfo {author} {\bibfnamefont {V.}~\bibnamefont
  {Biancalana}}, \bibinfo {author} {\bibfnamefont {Y.}~\bibnamefont
  {Dancheva}}, \bibinfo {author} {\bibfnamefont {T.}~\bibnamefont {Mansour}}, \
  and\ \bibinfo {author} {\bibfnamefont {L.}~\bibnamefont {Moi}},\ }\href
  {\doibase http://dx.doi.org/10.1063/1.3567410} {\bibfield  {journal}
  {\bibinfo  {journal} {J. Math. Phys.}\ }\textbf {\bibinfo {volume} {52}},\
  \bibinfo {eid} {033508} (\bibinfo {year} {2011})}\BibitemShut {NoStop}%
\bibitem [{\citenamefont {De~Nardis}\ \emph {et~al.}()\citenamefont
  {De~Nardis}, \citenamefont {Piroli},\ and\ \citenamefont
  {Caux}}]{DeNardisg2TimeEv}%
  \BibitemOpen
  \bibfield  {author} {\bibinfo {author} {\bibfnamefont {J.}~\bibnamefont
  {De~Nardis}}, \bibinfo {author} {\bibfnamefont {L.}~\bibnamefont {Piroli}}, \
  and\ \bibinfo {author} {\bibfnamefont {J.-S.}\ \bibnamefont {Caux}},\
  }\href@noop {} {\bibinfo  {journal} {arXiv:1505.03080}\ }\BibitemShut
  {NoStop}%
\bibitem [{\citenamefont {{De Nardis}}\ and\ \citenamefont
  {Caux}(2014)}]{2014_Nardis_JSTAT_P12012}%
  \BibitemOpen
\bibfield  {journal} {  }\bibfield  {author} {\bibinfo {author} {\bibfnamefont
  {J.}~\bibnamefont {{De Nardis}}}\ and\ \bibinfo {author} {\bibfnamefont
  {J.-S.}\ \bibnamefont {Caux}},\ }\href
  {http://stacks.iop.org/1742-5468/2014/i=12/a=P12012} {\bibfield  {journal}
  {\bibinfo  {journal} {J. Stat. Mech.: Th. Exp.}\ }\textbf {\bibinfo {volume}
  {2014}},\ \bibinfo {pages} {P12012} (\bibinfo {year} {2014})}\BibitemShut
  {NoStop}%
\bibitem [{\citenamefont {{Mehler}}(1866)}]{1866_Mehler_German}%
  \BibitemOpen
  \bibfield  {author} {\bibinfo {author} {\bibfnamefont {F.~G.}\ \bibnamefont
  {{Mehler}}},\ }\href {\doibase 10.1515/crll.1866.66.161} {\bibfield
  {journal} {\bibinfo  {journal} {{J. Reine Angew. Math.}}\ }\textbf {\bibinfo
  {volume} {66}},\ \bibinfo {pages} {161} (\bibinfo {year} {1866})}\BibitemShut
  {NoStop}%
\bibitem [{\citenamefont {Campbell}\ \emph {et~al.}(2015)\citenamefont
  {Campbell}, \citenamefont {Gangardt},\ and\ \citenamefont
  {Kheruntsyan}}]{2015_Campbell_PRL_114}%
  \BibitemOpen
  \bibfield  {author} {\bibinfo {author} {\bibfnamefont {A.~S.}\ \bibnamefont
  {Campbell}}, \bibinfo {author} {\bibfnamefont {D.~M.}\ \bibnamefont
  {Gangardt}}, \ and\ \bibinfo {author} {\bibfnamefont {K.~V.}\ \bibnamefont
  {Kheruntsyan}},\ }\href {\doibase 10.1103/PhysRevLett.114.125302} {\bibfield
  {journal} {\bibinfo  {journal} {Phys. Rev. Lett.}\ }\textbf {\bibinfo
  {volume} {114}},\ \bibinfo {pages} {125302} (\bibinfo {year}
  {2015})}\BibitemShut {NoStop}%
\bibitem [{\citenamefont {Kheruntsyan}\ \emph {et~al.}(2005)\citenamefont
  {Kheruntsyan}, \citenamefont {Gangardt}, \citenamefont {Drummond},\ and\
  \citenamefont {Shlyapnikov}}]{2005_Kheruntsyan_PRA_71}%
  \BibitemOpen
  \bibfield  {author} {\bibinfo {author} {\bibfnamefont {K.~V.}\ \bibnamefont
  {Kheruntsyan}}, \bibinfo {author} {\bibfnamefont {D.~M.}\ \bibnamefont
  {Gangardt}}, \bibinfo {author} {\bibfnamefont {P.~D.}\ \bibnamefont
  {Drummond}}, \ and\ \bibinfo {author} {\bibfnamefont {G.~V.}\ \bibnamefont
  {Shlyapnikov}},\ }\href {\doibase 10.1103/PhysRevA.71.053615} {\bibfield
  {journal} {\bibinfo  {journal} {Phys. Rev. A}\ }\textbf {\bibinfo {volume}
  {71}},\ \bibinfo {pages} {053615} (\bibinfo {year} {2005})}\BibitemShut
  {NoStop}%
\bibitem [{\citenamefont {van Amerongen}\ \emph {et~al.}(2008)\citenamefont
  {van Amerongen}, \citenamefont {van Es}, \citenamefont {Wicke}, \citenamefont
  {Kheruntsyan},\ and\ \citenamefont {van Druten}}]{2008_Amerongen_PRL_100}%
  \BibitemOpen
  \bibfield  {author} {\bibinfo {author} {\bibfnamefont {A.~H.}\ \bibnamefont
  {van Amerongen}}, \bibinfo {author} {\bibfnamefont {J.~J.~P.}\ \bibnamefont
  {van Es}}, \bibinfo {author} {\bibfnamefont {P.}~\bibnamefont {Wicke}},
  \bibinfo {author} {\bibfnamefont {K.~V.}\ \bibnamefont {Kheruntsyan}}, \ and\
  \bibinfo {author} {\bibfnamefont {N.~J.}\ \bibnamefont {van Druten}},\ }\href
  {\doibase 10.1103/PhysRevLett.100.090402} {\bibfield  {journal} {\bibinfo
  {journal} {Phys. Rev. Lett.}\ }\textbf {\bibinfo {volume} {100}},\ \bibinfo
  {pages} {090402} (\bibinfo {year} {2008})}\BibitemShut {NoStop}%
\bibitem [{\citenamefont {Armijo}\ \emph {et~al.}(2011)\citenamefont {Armijo},
  \citenamefont {Jacqmin}, \citenamefont {Kheruntsyan},\ and\ \citenamefont
  {Bouchoule}}]{2011_Armijo_PRA_83}%
  \BibitemOpen
  \bibfield  {author} {\bibinfo {author} {\bibfnamefont {J.}~\bibnamefont
  {Armijo}}, \bibinfo {author} {\bibfnamefont {T.}~\bibnamefont {Jacqmin}},
  \bibinfo {author} {\bibfnamefont {K.}~\bibnamefont {Kheruntsyan}}, \ and\
  \bibinfo {author} {\bibfnamefont {I.}~\bibnamefont {Bouchoule}},\ }\href
  {\doibase 10.1103/PhysRevA.83.021605} {\bibfield  {journal} {\bibinfo
  {journal} {Phys. Rev. A}\ }\textbf {\bibinfo {volume} {83}},\ \bibinfo
  {pages} {021605} (\bibinfo {year} {2011})}\BibitemShut {NoStop}%
\bibitem [{\citenamefont {Jacqmin}\ \emph {et~al.}(2011)\citenamefont
  {Jacqmin}, \citenamefont {Armijo}, \citenamefont {Berrada}, \citenamefont
  {Kheruntsyan},\ and\ \citenamefont {Bouchoule}}]{2011_Jacqmin_PRL_106}%
  \BibitemOpen
  \bibfield  {author} {\bibinfo {author} {\bibfnamefont {T.}~\bibnamefont
  {Jacqmin}}, \bibinfo {author} {\bibfnamefont {J.}~\bibnamefont {Armijo}},
  \bibinfo {author} {\bibfnamefont {T.}~\bibnamefont {Berrada}}, \bibinfo
  {author} {\bibfnamefont {K.~V.}\ \bibnamefont {Kheruntsyan}}, \ and\ \bibinfo
  {author} {\bibfnamefont {I.}~\bibnamefont {Bouchoule}},\ }\href {\doibase
  10.1103/PhysRevLett.106.230405} {\bibfield  {journal} {\bibinfo  {journal}
  {Phys. Rev. Lett.}\ }\textbf {\bibinfo {volume} {106}},\ \bibinfo {pages}
  {230405} (\bibinfo {year} {2011})}\BibitemShut {NoStop}%
\end{thebibliography}%

\clearpage

\section*{Supplemental Material}

\subsection{Matrix elements and initial state on the ring}
 The initial state after the Bragg pulse is easily obtained from the matrix elements of the operator
  \begin{equation}
  \label{eq:1}
  \hat{U}_B(q,A)  = e^{-iA\int dx \cos(q x)\hPsi^{\dagger}(x)\hPsi(x)}.
  \end{equation}
  In the  Tonks-Girardeau (TG) limit on a ring geometry we use the eigenstates
\begin{equation} \label{eq:quantum_state_definition_LL}
\ket{\blam} =  \frac1{\sqrt{N!}}\int_0^L d^{N}x \, \psi_{N}(\boldx|\blam)\,\hPsi^{\dagger}(x_{1})\dots\hPsi^{\dagger}(x_{N})\,|0\rangle \epc
\end{equation}
with wavefunctions given by
\begin{equation}
\psi_{N}(\boldx|\blam) = \frac1{\sqrt{N!}}\,\text{det} 
\left[ e^{ix_{l} \lam_j} \right]  \displaystyle{\prod_{1\leq l<j \leq N}} \text{sgn}(x_{j}-x_{l}) \epp
\end{equation}
 By commuting $\hat{U}_B(q,A) $ through the creation  operators, the
 matrix elements can be expressed as
  \begin{align}
  \label{eq:9}
    \bra{\blam}  \hat{U}_B(q,A) \ket{\bmu}  
    = \frac{1}{N!} \int d^Nx d^Nx'
    \psi_{N}(\boldx|\blam)^{*}\psi_{N}(\boldx'|\bmu) \notag \\
   \times \;  e^{-iA\sum_n \cos(qx_n)}  
    \bra{0} \prod_n\hPsi(x'_n) \prod_j \hPsi^{\dag}(x_j)\ket{0}.
  \end{align}
The expectation value of the bosonic operators conspires with the
signs in the Tonks-Girardeau wavefunctions leading to a determinant
of $\delta$-functions. Treating the coordinates as dummy variables under the
integral sign, it is easy to rewrite the integral in factorized form
as a determinant of integrals of the form
\begin{equation}
\label{eq:2}
 \frac{1}{L} \int_0^L dx e^{ix (\lambda_j - \mu_k) - iA \cos( q x)  } = I_{\frac{\lam_j-\mu_k}{q}}(-iA)\, \delta^{(q)}_{\lam_j,\mu_k},
\end{equation}
where we define $\delta^{(q)}_{\lambda,\mu} = \delta_{(\lambda -\mu)\,
  \text{mod}\,q,0}$ and where we used that $\lambda_j-\mu_k$ and $q$ lie on the momentum lattice $(2\pi \alpha/L)$ with $\alpha \in \mathbb{Z}$.
This results in the matrix elements
\begin{equation}
\label{eq:3}
 \frac{\bra{\bmu}{\hU_B (A)}\ket{\blam}}{L^N} = \, \text{det}_N\! \left[ \left( I_{\frac{\lam_j-\mu_k}{q}}(-iA)\, \delta^{(q)}_{\lam_j,\mu_k} \right)_{j,k} \right] .
\end{equation}
The initial state
\begin{equation}
\label{eq:4}
 \ket{\psi_{q,A}} = \hat{U}_B(q,A) \ket{\psi_{GS}}
\end{equation}
is easily expressed in the Tonks-Girardeau eigenbasis using the matrix
elements $\bra{\bmu} \hat{U}_B(q,A) \ket{\psi_{GS}}$.

\subsection{The stationary state on a ring from a GGE and the Quench Action approach}
In order to implement the GGE logic~\cite{2007_Rigol_PRL_98,2008_Rigol_NATURE_452}, one starts with computing the conserved charges on the initial state. Let us focus on the case $q >2 \lam_F$, for which the overlaps $\langle \blam | \psi_{q,A} \rangle$ coming from Eq.~\eqref{eq:3} reduce to a simple product of $N$ modified Bessel functions. While odd charges are trivially zero, for the even charges we find at finite system size
\begin{align}
& \quad\, \smatrixel{\psi_{q,A}}{\hat{Q}_{2\alpha}}{\psi_{q,A}} \notag \\
&= \sum_{j=1}^N \sum_{\beta\in\mathbb{Z}} \left|  I_{\beta}(iA)\right|^2 \big( \lam_j (\beta) \big)^{2\alpha} \notag \\
&=  \sum_{j=1}^N \sum_{\beta\in\mathbb{Z}} \left|  I_{\beta}(iA)\right|^2 \sum_{l=0}^{\alpha} {2\alpha \choose 2l} \big(\lam_{j}^\GS\big)^{2(\alpha-l)} (q\beta)^{2l} \raisetag{.5cm} \notag\\
&=  \sum_{j=1}^N \sum_{l=0}^{\alpha} {2\alpha \choose 2l} \big(\lam_{j}^\GS\big)^{2(\alpha-l)} q^{2l} B_{2l,0}(A) \epc
\end{align}
where we defined $\lam_j (\beta) = \lam_{j}^\GS + q \beta $ and where the coefficients $B_{2l,0}$ come from the sum over the order of the Bessel functions and are known recursively \cite{2011_Bevilacqua_JMP_52}. The sum over particles $j$ can be performed, after which the thermodynamic limit can be taken,
\begin{align} \label{eq:expecation_value_local_charge_large_q}
& \quad \, \limth \smatrixel{\psi_{q,A}}{\hat{Q}_{2\alpha}/N}{\psi_{q,A}} \notag \\
& = \sum_{l=0}^{\alpha} {2\alpha \choose 2l} \frac{(n \pi)^{2(\alpha-l)}q^{2l}}{2(\alpha-l)+1} B_{2l,0}(A) \epc
\end{align}
where $n$ is the average particle density. For example, the energy density pumped into the system by an instantaneous Bragg pulse is given by
\begin{equation}
\limth \big( \smatrixel{\psi_{q,A}}{\hat{Q}_{2}/N}{\psi_{q,A}} - \smatrixel{\psi_\GS}{\hat{Q}_{2}/N}{\psi_\GS} \big)  = \frac{q^2 A^2}{2} \epp
\end{equation}
One can show that the saddle-point density 
\begin{equation} \label{eq:supmat_saddle_point_bragg_pulse}
\rhosp_{q,A}(\lam) = \frac{1}{2\pi} \! \sum_{\beta\in\mathbb{Z}} \big[ \theta(\lam-\beta q + \lam_F) - \theta(\lam-\beta q - \lam_F) \big] \!  \left| I_{\beta} ( iA) \right|^2 
\end{equation}
reproduces these values of the conserved charges, i.e. $L \int_{-\infty}^\infty d\lam\, \rho^{\text{sp}}_{q,A}(\lam) \, \lam^{2\alpha} = \limth \smatrixel{\psi_{q,A}}{\hat{Q}_{2\alpha}}{\psi_{q,A}}$ for all $\alpha \in\mathbb{N}$, by performing the integral and recasting the infinite sum into the coefficients $B_{2a,0}$. Since the local conserved charges (if well defined) uniquely determine the saddle point, we have thus found the saddle-point density after a Bragg pulse for $q >2 \lam_F$. For smaller Bragg momenta $q <2 \lam_F$ the computation becomes considerably more difficult due to the determinant structure of the overlaps, but one can show that the saddle-point density given in Eq.~\eqref{eq:supmat_saddle_point_bragg_pulse} is still correct.

The Quench Action (QA) approach~\cite{2013_Caux_PRL_110,2014_DeNardis_PRA_89} reproduces this saddle-point density for $q >2 \lam_F$. As a consequence of working in the Tonks-Girardeau regime, there are many microstates with exactly the same overlap. We can rephrase the overlaps as
\begin{equation}
\langle \{ \lam_j(\beta_j) \}_{j=1}^N | \psi_{q,A} \rangle = L^N \prod_{\alpha=-\infty}^\infty  \left[I_{\alpha} (-iA) \right]^{n_\alpha} \epc
\end{equation}
where $n_\alpha$ is the number of rapidities $j$ with $\beta_j=\alpha$ and $\alpha \in \mathbb{Z}$. In the thermodynamic limit these numbers are given by
\begin{equation}
n_\alpha = L \int_{\alpha q - \lam_F}^{\alpha q + \lam_F} d\lam \, \rho(\lam) \epp
\end{equation}
The normalized overlap coefficients $S_{\{ \beta_j \}} = - \ln \big( \langle \{ \lam_j(\beta_j) \}_{j=1}^N | \psi_{q,A} \rangle/L^N \big)$ have a well-defined thermodynamic limit,
\begin{align}
S [ \rho] &=  \limth \text{Re}\, S_{\{ \beta_j \}} \\ 
&= - L \sum_{\alpha=-\infty}^\infty \int_{\alpha q - \lam_F}^{\alpha q + \lam_F} d\lam \, \rho(\lam) \ln \left[ \left| I_{\alpha} (-iA) \right| \right]  \\
&=  L \! \int_{-\infty}^\infty \!\!\! d\lam \, \rho(\lam) \!\!  \sum_{\alpha=-\infty}^\infty \!\! \big[ \theta(\lam - \alpha q - \lam_F) \notag \\
& \qquad \qquad - \theta(\lam - \alpha q + \lam_F) \big] \log \left[ \left| I_{\alpha} (iA) \right| \right], 
\end{align}
where $\theta$ is the Heaviside step function and we used that $|I_n(-z)|=|I_n(z)|$. Furthermore, in the thermodynamic limit $\lam_F= \pi n$, where $n$ is the average particle density. The noncontinuous integrand will serve as the driving term of the GTBA equations. Note that in the second line we implicitely assume that $\rho(\lam)=0$ when $\lam \notin \left[ \lam - \alpha q - \lam_F, \lam - \alpha q + \lam_F \right]$ for any $\alpha \in \mathbb{Z}$. The reason is that for Bethe states that do not obey this condition, the overlap is exactly zero (rapidities will never end up in those regions) and therefore $S[\rho]=\infty$. These states are therefore infinitely suppressed in the Quench Action saddle-point equations. Another way of seeing this is that originally the functional integral in the quench action approach is a sum over states with non-zero overlaps and these states are not in that sum.

Even when you restrict the support of the density function to these intervals, in this ensemble of states there are still many microstates that have zero overlap with the Bragg-pulsed ground state. The reason is that when a rapidity $\lam_j^\GS$ has moved to an interval $\alpha=\beta_j$, it is not in the other intervals $\alpha \neq\beta_j$ and therefore leaves a hole there. This alters the usual form of the Yang-Yang entropy significantly. Given the fillings $\{ n_\alpha \}_{\alpha=-\infty}^\infty$, the finite-size entropy is
\begin{subequations}
\begin{align}
e^{S_{\text{YY},\{ n_\alpha\}}} \! 
&=  \frac{N!}{\prod_{\alpha=-\infty}^\infty (n_\alpha !)} \epc
\end{align}
\end{subequations}
which leads to a modified Yang-Yang entropy for the Bragg pulse from the ground state,
\begin{equation}
S_{\text{YY}}[\rho] = - L \int_{-\infty}^\infty d\lam \, \rho(\lam) \log [2\pi\rho(\lam)] \epc
\end{equation}
where we used the Tonks-Girardeau Bethe equation $2\pi [\rho(\lam) + \rho_h(\lam)]=1$. The variation of the quench action should be restricted to densities for which $\rho(\lam)=0$ when $\lam \notin \left[ \lam - \alpha q - \lam_F, \lam - \alpha q + \lam_F \right]$ for any $\alpha \in \mathbb{Z}$. Also, a Lagrange multiplier $h$ is added to fix the particle density to $n=N/L$. The resulting GTBA equation is not an integral equation because of the Tonks-Girardeau limit,
\begin{align}
&0=2 \!\! \sum_{\alpha=-\infty}^\infty \!\!\! \left[ \theta(\lam \! - \! \alpha q \! - \! \lam_F) \! - \! \theta(\lam \! - \! \alpha q \! + \! \lam_F) \right] \log \left[ \left| I_{\alpha} (iA) \right| \right] \notag \\
&  \qquad  \qquad \qquad \qquad + \log\! \left( 2\pi\rhosp_{q,A}(\lam)\right) + 1 - h  \epp
\end{align}
This is solved by the normalized saddle-point density of Eq.~\eqref{eq:supmat_saddle_point_bragg_pulse}, with $h=1$. For smaller Bragg momenta $q <2 \lam_F$ the derivation of the saddle-point distribution using the QA approach remains an open problem, since the determinant structure of the overlaps prevents obtaining a straightforward thermodynamic limit of the overlap coefficients that is expressible in terms of a root density $\rho(\lam)$. 

Moreover, one could question whether the time evolution of simple observables obtained from the QA approach using the saddle-point density of Eq.~\eqref{eq:supmat_saddle_point_bragg_pulse} is valid also for small Bragg momenta $q < 2\lam_F$. However, the analysis of the FB mapping does not show any qualitative differences for Bragg momenta smaller or bigger than $2 \lam_F$ and the agreement with the time evolution of the QA approach is excellent for all $q>0$. It therefore seems safe to assume that the time evolution from the QA approach is valid for all Bragg momenta.

\subsection{Time evolution of the momentum distribution function on a ring in the thermodynamic limit}

In Ref.~\cite{2014_Nardis_JSTAT_P12012} the thermodynamic limit of the matrix elements of the the one-body density matrix between states with a countable number $n_{\rm e}$ of particle-hole excitations $\{  h_j \to p_j \}_{j=1}^{n_{\rm e}}$ on  a thermodynamic state $| \rho \rangle$ was computed. The result is obtained by decomposing the elements into a Fredholm determinant and a finite-size determinant accounting for the excitations. The result is
\begin{align}
&\matrixel{\rho}{\hPsi^\dag(x) \hPsi(0)}{\rho,\{ h_j \to p_j \}_{j=1}^{n_{\rm e}}} \nonumber \\
& = L^{-n_{\rm e}} e^{i\frac{x}{2} \sum_{j=1}^{n_{\rm e}} (p_j-h_j)} \Big\{ \text{Det}(1+ K'\rho) \det_{i,j=1}^{n_{\rm e}} \left[W'\big(h_i,p_j\big)\right] \nonumber \\&
 - \text{Det}(1+K\rho) \det_{i,j=1}^{n_{\rm e}} \left[W\big(h_i,p_j\big)\right] \Big\} , 
\end{align}
where $\rho(\lambda)$ is the density of rapidities of the thermodynamic state. 
The Fredholm determinants are denoted by $\text{Det}$, where $(K\rho)(\lam,\mu) = K(\lam,\mu)\rho(\mu)$. The kernels are given by
\begin{align}
& K'(\lam,\mu) = K(\lam,\mu) + n e^{-i\frac{x}{2}(\lam+\mu)} \epc \\&  K(\lam,\mu) = -4 n \frac{\sin\left(\frac{x}{2}(\lam-\mu)\right)}{\lam-\mu} \epc
\end{align}
where $n=N/L$ is the density. The function $W$ is defined as
\begin{align}
W(\lam,\mu) & =  \left( (1+K\rho)^{-1} K \right) (\lam,\mu) \epc \\
W'(\lam,\mu) & =  \left( (1+K'\rho)^{-1} K' \right) (\lam,\mu) \epc
\end{align}
or equivalently via the integral equations
\begin{align}
& W(\lam,\mu) + \int_{-\infty}^\infty d\nu\, K(\lam,\nu) \rho(\nu) W(\nu,\mu) = K(\lam,\mu) \epc \\&
W'(\lam,\mu) + \int_{-\infty}^\infty d\nu\, K'(\lam,\nu) \rho(\nu) W'(\nu,\mu) = K'(\lam,\mu) \epp
\end{align}
The QA approach yields the following expression for the time evolution of the one-body density matrix in the thermodynamic limit 
\begin{align}\label{eq:sum_exc_QA}
& \matrixel{\psi_{q,A}(t)}{\hPsi^\dag(x) \hPsi(0)}{\psi_{q,A}(t)} \nonumber \\& = \Re \sum_{n_{\rm e}=0}^{\infty} \frac{1}{(n_{\rm e}!)^2} \left( \prod_{j=1}^{n_{\rm e}} L^2 \int_{-\infty}^{\infty} dh_j  dp_j \:   \varphi_{-}^{(t)}(h_j) \varphi^{(t)}_{+}(p_j)  \right)\nonumber  \\& \times   \matrixel{\rho_{q,A}}{\hPsi^\dag(x) \hPsi(0)}{\rho_{q,A},\{ h_j \to p_j \}_{j=1}^{n_{\rm e}}} \epc
\end{align}
where the effective densities of holes and particles are given by
\begin{align}
&\varphi_{-}^{(t)}(h_j) = e^{ \delta s(h_j) + i \delta \omega(h_j)t} \rho_{q,A}(h_j) \epc \\
& \varphi_{+}^{(t)}(p_j) = e^{- \delta s(h_j) - i \delta \omega(h_j)t} \rho^h_{q,A}(p_j) \epc
\end{align}
with the density of holes of the saddle point given by $\rho^h_{q,A}(p) = \frac{1}{2 \pi} - \rho_{q,A}(p) $.
The differential overlap for a single particle-hole $e^{- \delta s(p) + \delta s (h)}$ is obtained by taking a finite-size realization of the saddle point state $| \boldsymbol{\lambda}_{q,A} \rangle \to | \rho_{q,A }\rangle $ and the modified state obtained by performing a single particle-hole on $| \boldsymbol{\lambda}_{q,A} \rangle $. The ratio of the two overlaps in the thermodynamic limit gives the differential overlap
\begin{equation}
e^{- \delta s(p) + \delta s (h)} = \lim_{N \to \infty} \frac{\langle \psi_{q,A} | \boldsymbol{\lambda}_{q,A} , h \to p \rangle }{\langle \psi_{q,A} | \boldsymbol{\lambda}_{q,A} \rangle}  \epp
\end{equation}
The same argument gives the energy of the single particle-hole excitations
\begin{equation}
e^{- i \delta \omega(p)t + i \delta \omega (h)t} = e^{- i  p^2 t + i  h^2 t } \epp
\end{equation}
Since the overlaps only couple states such that the difference of their rapidities are multiples of $q$, the sum over the excitations on the saddle point state reduces to 
\begin{align}
& \frac{1}{(n_{\rm e}!)^2}\prod_{j=1}^{n_{\rm e}} L^2 \int_{-\infty}^{\infty} dh_j  dp_j \:   \varphi_{-}^{(t)}(h_j) \varphi^{(t)}_{+}(p_j) \nonumber \\
& \to  \frac{1}{n_{\rm e}!}\prod_{j=1}^{n_{\rm e}} L \int_{-\infty}^{\infty} dh_j  \sum_{\beta_j \in \mathbb{Z}}   \varphi_{-}^{(t)}(h_j) \frac{\varphi^{(t)}_{+}(h_j + \beta_j q)}{\rho^h_{q,A}(h_j + \beta_j q) } \epp
\end{align} 
After this substitution the sum in Eq.~\eqref{eq:sum_exc_QA} becomes the definition of a Fredholm determinant, and using the definition of the saddle-point distribution in Eq.~\eqref{eq:supmat_saddle_point_bragg_pulse},  the expression for the time evolution of the one-body density matrix can be rewritten as the difference of two Fredholm determinants of two infinite block matrices where each block $S_{\alpha, \beta}$ and $S'_{\alpha, \beta}$ for any $\alpha,\beta \in \mathbb{Z}$ is an operator acting on $\Lambda = [-\lambda_F, \lambda_F]$,
\begin{align}\label{eq:TE1}
  \langle  \Psi_{q,A}(t)  | &  \Psi^\dagger(x)  \Psi(0)  |\Psi_{q,A}(t)  \rangle \nonumber \\= 
\Re \Big[  \text{Det}_\Lambda&  \left(\mathbf{1} \delta_{\alpha, \beta}+  S'_{\alpha, \beta}      \right)_{\alpha,\beta \in \mathbb{Z}} \nonumber \\&  -  
\text{Det}_\Lambda \left(\mathbf{1} \delta_{\alpha, \beta}+   S_{\alpha, \beta}   \right)_{\alpha,\beta \in \mathbb{Z}}  \Big]\epc
\end{align}
with the operators given by
\begin{equation}
 S'_{\alpha, \beta}  (u,v)=  \sum_{\gamma \in \mathbb{Z}} 
   \zeta^{(t)}_\gamma(u + \alpha q) K' (u + \alpha q, v + (\beta + \gamma)  q)  \Phi^{(t)}_{\beta, \gamma} \epc \nonumber
\end{equation}
\begin{equation}
 S_{\alpha, \beta}  (u,v)=  \sum_{\gamma \in \mathbb{Z}} 
   \zeta^{(t)}_\gamma(u + \alpha q) K (u + \alpha q, v + (\beta + \gamma) q)  \Phi^{(t)}_{\beta, \gamma} \epp
\end{equation}
Here $u,v \in [-\lambda_F, \lambda_F ]$ and $\mathbf{1}$ is the identity operator. The coefficients $\Phi^{(t)}_{\beta, \gamma}$ and the function $\zeta_\gamma^{(t)}(u)$ are given by
\begin{equation}
\Phi^{(t)}_{\beta, \gamma} = \frac{ I_{\beta}( i A) I_{\beta +  \gamma} (-i A)}{2 \pi}    e^{- it (q \gamma)^2  + i x q \gamma/2} \epc
\end{equation}
\begin{equation}
\zeta^{(t)}_\gamma(u)= e^{-2 i t q \gamma u } \epp
\end{equation}
In order to obtain the time evolution of the momentum distribution $ \hat{n}(k,t)$ one needs to restrict the sum in Eq.~\eqref{eq:sum_exc_QA} to excitations with zero total momentum, namely
\begin{align}\label{eq:sum_exc_QA_2}
&  \hat{n}(k,t) = \nonumber \\&  \text{FT}   \Big\{ \sum_{n_{\rm e}=0}^{\infty} \frac{1}{n_{\rm e}!} \left( \prod_{j=1}^{n_{\rm e}} \int_{-\infty}^{\infty} \!\! dh_j  \! \sum_{\beta_j \in \mathbb{Z}}\:   \varphi_{-}^{(t)}(h_j) \frac{\varphi^{(t)}_{+}(h_j + \beta_j q)}{\rho^h_{q,A}(h_j + \beta_j q) }  \right)\nonumber  \\& \times   L^{n_{\rm e}}\matrixel{\rho_{q,A}}{\Psi^\dag(x) \Psi(0)}{\rho_{q,A},\{ h_j \to h_j + q \beta_j \}_{j=1}^{n_{\rm e}}} \nonumber \\& \times   \delta_{\sum_{j=1}^{n_{\rm e}} \beta_j ,0}  \Big\} 
\end{align}
where we denoted the Fourier transform as $\text{FT} \{  f(x) \} = \int_{-\infty}^{\infty} dx \: f(x) e^{- i k x}  $. Using the identity 
\begin{equation}
\int_{-\pi}^{\pi}\frac{d v}{2 \pi} e^{- i \beta v} = \delta_{\beta,0} \epc
\end{equation}
we obtain 
 \begin{widetext}
  \begin{equation}
  \hat{n}(k,t) \nonumber =  \text{FT}   \Big\{ \Re \int_{-\pi}^{\pi} \frac{d\kappa}{2 \pi} \left[
\text{Det}_\Lambda \left(\mathbf{1} \delta_{\alpha, \beta}+  S^{(\kappa)}{}'_{\alpha, \beta}       \right)_{\alpha,\beta \in \mathbb{Z}}    -  
\text{Det}_\Lambda \left(\mathbf{1} \delta_{\alpha, \beta}+   S^{(\kappa)}{}_{\alpha, \beta}         \right)_{\alpha,\beta \in \mathbb{Z}}   \right]\Big\} \epc
\end{equation}
\begin{equation}
 S^{(\kappa)}{}'_{\alpha, \beta}  (u,v)=  \sum_{\gamma \in \mathbb{Z}} 
   \zeta^{(t)}_\gamma(u + \alpha q) K' (u + \alpha q, v + (\beta + \gamma)  q)  \Phi^{(t)}_{\beta, \gamma} e^{-i \kappa \gamma} \epc \nonumber
\end{equation}
\begin{equation}
S^{(\kappa)}_{\alpha, \beta}  (u,v)=  \sum_{\gamma \in \mathbb{Z}} 
   \zeta^{(t)}_\gamma(u + \alpha q) K (u + \alpha q, v + (\beta + \gamma) q)  \Phi^{(t)}_{\beta, \gamma}  e^{-i \kappa \gamma} \epp
\end{equation}
\end{widetext}

\subsection{Time-evolved single-particle states in the trap}
The propagator for the quantum harmonic oscillator (Mehler kernel) is given by
\begin{align}
K(x,y;t) = &\sqrt{\frac{m\omega}{2\pi i \sin (\omega t)}} \times \notag \\
&\times \exp\left(\frac{- m \omega (x^2 + y^2 ) \cos(\omega t) + 2 m \omega xy}{2 i \sin(\omega t)}\right).
\end{align} 
The single particle (SP) wavefunctions after the Bragg pulse can then be time evolved by integrating the initial wavefunctions including the cosine phase with the propagator:
\begin{align}
\psi_j(x;t) = \int_{-\infty}^{\infty} &dy  K(x,y;t) e^{-iA \cos(qx)}\psi_j(y) \notag \\
=\sum_{\beta=-\infty}^{\infty}&I_{\beta}(-iA)e^{-i\beta q \cos(\omega t)\left(x+\frac{\beta q}{2m\omega}\sin(\omega t)\right)} \notag \\
 &\;\psi_j(x+ \tfrac{\beta q}{m\omega}\sin(\omega t))  e^{-i \omega(j+\frac{1}{2})t} \epc
\label{eq:supp_harm_phit}
\end{align}
where $\psi_j(x)$ are the groundstate harmonic eigenfunctions 
\begin{align}
\psi_j(x) = \frac{1}{\sqrt{2^j j!}} \left(\frac{m \omega}{\pi}\right)^{1/4} e^{-\frac{m\omega x^2}{2}} H_j\left(\sqrt{m \omega} x\right).
\label{eq:supmat_harm_eig}
\end{align}
The result in Eq. \eqref{eq:supp_harm_phit} has been obtained by using the following two identities
\begin{align}
&e^{-i z \cos(\phi)} = \sum_{n=-\infty}^{\infty}  I_n(- i z) e^{-i n \phi} \epc \\
&\int_{-\infty}^{\infty} \der x e^{-(x-y)^2} H_j(\alpha x) 
= \sqrt{\pi} (1-\alpha^2)^{\sfrac{j}{2}} H_j\left(\frac{\alpha y}{\sqrt{1-\alpha^2}}\right).
\end{align}

\subsection{Exact momentum distribution at $t=0$ for arbitrary interactions}

The one-body density matrix at $t=0$ (after the Bragg pulse) is given by
\begin{multline}
\label{eq:30}
  \langle \hat{U}_B^{\dag} (q,A) \hPsi^{\dag}(x)\hPsi(y) \hat{U}_B (q,A)
  \rangle \\
  = \langle \hPsi^{\dag}(x)\hPsi(y)\rangle e^{-i 2 A \sin\left(q \frac{x-y}{2}\right)\sin\left(q \frac{x+y}{2} \right)}.
\end{multline}
The latter equality follows strictly from the commutation relations of the Bose
fields with the density and thus holds irrespective of interaction or
geometry. 
For the case of the ring geometry, the associated momentum distribution function (MDF) is
\begin{equation}
\label{eq:32}
 \langle \hat{n}(k,t=0)\rangle
  = \frac{1}{L}\int_0^L d\xi e^{ik\xi}I_0\left(i 2A \sin(q\xi/2)\right)\langle \hPsi^{\dag}(\xi)\hPsi(0)\rangle \epc
\end{equation}
where we defined $\xi = x-y$ and used the integral
\begin{multline}
\label{eq:51}
 \frac{1}{L}\int_0^L dy e^{-i 2A\sin(q \xi/2)\sin(qy+q\xi/2) } \\
 = I_0\left(i 2A \sin(q \xi/2)\right) \epc
\end{multline}
 under the assumption that $qL/2\pi$ is integer.
\begin{figure}[ht]
	\includegraphics[width=0.99 \columnwidth]{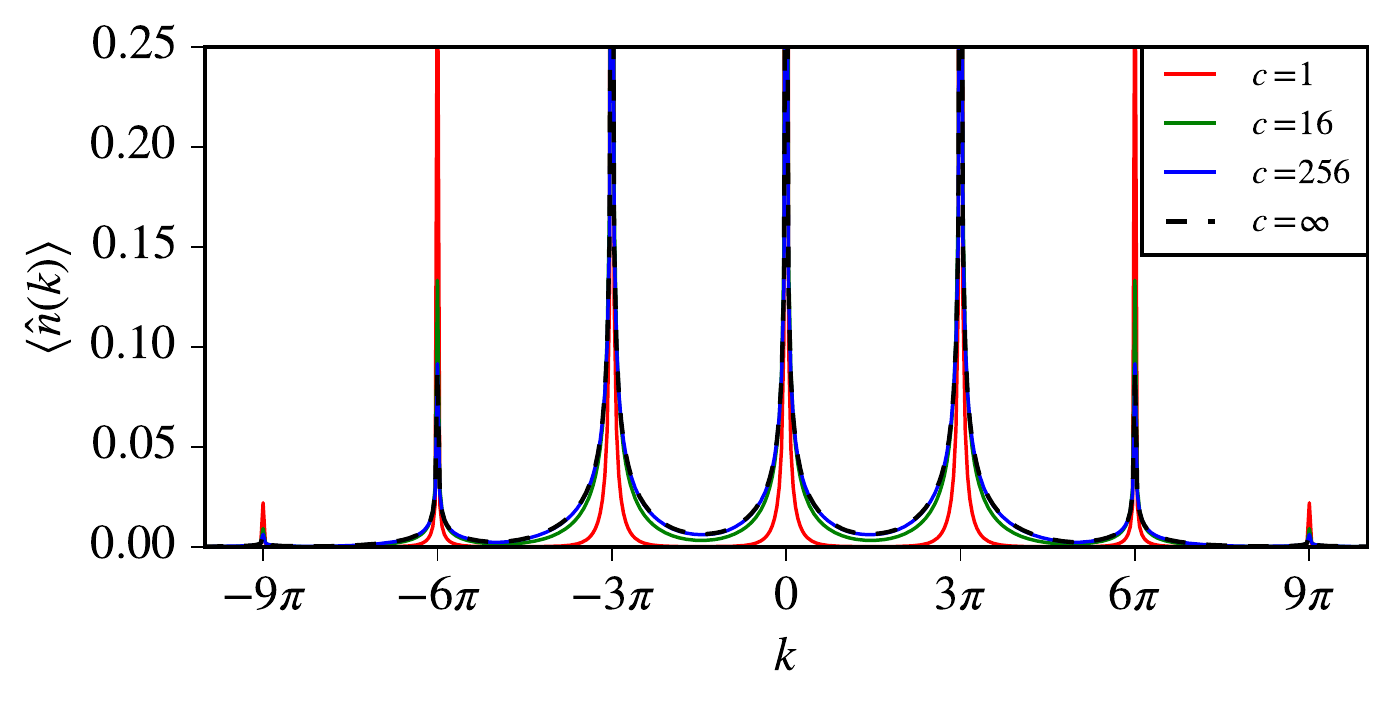}
		\caption{The initial MDF ($t = 0$) for $A = 1.5$, $q=3\pi$  and different values of the interaction strength $c$. The finite-$c$ interactions cause a decrease of the width of the satellites but do not influence their relative heights.}
	\label{fig:t0mom}
\end{figure}
Using the convolution theorem we obtain
\begin{equation}
\label{eq:53}
 \langle \hat{n}(k,t=0)\rangle =\sum_{k'} f(k')\langle \hat{n}(k-k')\rangle_{\rm GS}
\end{equation}
where
\begin{equation}
\label{eq:54}
 f(k) = \frac{1}{L}\int_0^L dx e^{ikx}I_0(i 2 A \sin(q x/2))
\end{equation}
and $\langle \hat{n}(k)\rangle_{\rm GS}$ is the MDF of the ground state.

Using the expansion $ I_0 (z) = \sum_{n=0}^{\infty} (\frac{1}{4}z^2)^n/(n!)^2$
one finds
\begin{multline}
\label{eq:56}
 I_0(i 2 A \sin(q x/2)) =\sum_{n=0}^{\infty} \frac{ (-1)^n}{(n!)^2}A^{2n}\sin^{2n}(qx/2)\\
 =\sum_{n=0}^{\infty}\sum_{l=-n}^n \frac{(-1)^{n+l}(2n)!}{(n!)^2(n-l)!(n+l)!}  \left(\frac{A}{2}\right)^{2n}e^{ilqx} \epc
\end{multline}
where we used the binomium to expand  in plane waves.

The order of the sums can now be interchanged. Defining the coefficients
\begin{equation}
\label{eq:57}
 c_l(A) = \sum_{n=|l|}^{\infty} \frac{(-1)^{n+l}(2n)!}{(n!)^2(n-l)!(n+l)!}  \left(\frac{A}{2}\right)^{2n}
\end{equation}
 we obtain $f(k) = \sum_l c_l \delta_{k,lq}$. The coefficients
 $c_l(A)$ can in fact be resummed and expressed in terms of a
hypergeometric function
\begin{equation}
\label{eq:59}
 c_l (A)=\frac{A^{2|l|}}{(|l|!)^2 2^{2|l|}}\tensor*[_{1}]{\mathrm{F}}{_{2}}\left( \frac{2|l|+1}{2};|l|+1,2|l|+1;-A^2 \right).
\end{equation}
The  $t=0$ post-pulse MDF can therefore
be exactly expressed in terms of the MDF before the
pulse $\langle \hat{n}(k)\rangle_{\rm {GS}}$ as
\begin{equation}
\label{eq:nkt0}
  \langle \hat{n}(k,t=0) \rangle = \sum_{l=-\infty}^{\infty}c_l(A)\langle \hat{n}(k+ l q)\rangle_{\rm {GS}}\;.
\end{equation}
Note that this result holds for arbitrary interaction strength $c$ with $\langle \hat{n}(k)\rangle_{\rm {GS}}$ the appropriate ground state MDF.
The result is plotted in Fig.~\ref{fig:t0mom} for different values of $c$. 
The influence of the finite interactions resides solely in the groundstate MDF $\langle \hat{n}(k + ql) \rangle_{\text{GS}}$, leading to a decreasing width of the peaks as one goes from the hard-core limit ($c\rightarrow \infty$) to the BEC limit ($c\rightarrow0$). In contrast, Eq. \eqref{eq:nkt0} shows that their relative heights are completely determined by the value of $A$.

\subsection{Local density approximation}

The local density approximation (LDA) for the gas in a parabolic trap
amounts to replacing the value for the mean density in  the Quench Action
result for the short distance fluctuations with a space-dependent
density profile corresponding to the ground state in the
trap. This result is considerably improved when one introduces the
classical harmonic motion of the density profile in accordance with
the exact $t=0$ MDF Eq.~\eqref{eq:nkt0}.

In the thermodynamic limit the ground state density profile in a harmonic trap is given by
\begin{equation}
\label{eq:5}
 \rho_{\mathrm{GS}}(x) = \langle \hat{\rho}(x)\rangle_{\mathrm{GS}} =\frac{1}{\pi}\sqrt{m N \omega - m^2\omega^2 x^2}.
\end{equation}
The QA result for the time-evolved density profile yields
\begin{align} 
 \rho_{\rm QA}(x,t;n) =& \limth \matrixel{ \psi_{q,A} (t)}{ \hat{\rho}(x)  }{ \psi_{q,A} (t) }   \notag \\
 =&  \frac{n m}{ q  \lam_F t} 
 \sum_{\beta=-\infty}^{\infty} J_{\beta }(-2A\sin(q^2\beta t/2m)) \times \notag \\
\quad &\cos(xq\beta)  \frac{\sin(q \lam_F  \beta t/m )}{\beta } \epc
 \label{eq:supmat_time_evolution_density}
\end{align}
with  $n$ the mean density on the ring. The result for the LDA in the trap then reads  
\begin{multline}
\label{eq:10}
 \rho_{\rm LDA}(x,t) =\\ \sum_l c_l(A)\;\rho_{\rm QA}\left(x -  \frac{l q}{\omega m}\sin(\omega t),t; \rho_{\rm GS}(x- \frac{l q}{\omega m}\sin(\omega t))\right) \epc
\end{multline}
where the coefficients $c_l(A)$ are given in Eq. \eqref{eq:59}.

\end{document}